\documentclass[journal]{IEEEtran}
\IEEEoverridecommandlockouts
\usepackage{cite}
\usepackage{comment}
\usepackage{amsmath,amssymb,amsfonts}
\usepackage{amsmath}
\usepackage{breq n}
\usepackage{algorithmic}
\usepackage{graphicx}
\usepackage{textcomp}
\usepackage{xcolor}
\usepackage{verbatim} 
\usepackage{balance}
\usepackage{tikz}

\newtheorem{lemma}{\textbf{Lemma}}
\newtheorem{prop}{\textbf{Proposition}}

\def\BibTeX{{\rm B\kern-.05em{\sc i\kern-.025em b}\kern-.08em
    T\kern-.1667em\lower.7ex\hbox{E}\kern-.125emX}}

\begin{document}
\newcommand{\blackline}{\raisebox{2pt}{\tikz{\draw[-,black!40!black,solid,line width = 0.9pt](0,0) -- (5mm,0);}}}
\newcommand{\blueline}{\raisebox{2pt}{\tikz{\draw[-,black!40!blue,solid,line width = 0.9pt](0,0) -- (5mm,0);}}}

\newcommand{\rectangleblack}{\raisebox{0pt}{\tikz{\draw[black,solid,line width = 1.0pt](2.mm,0) rectangle (3.5mm,1.5mm);\draw[-,black,solid,line width = 1.0pt](0.,0.8mm) -- (5.5mm,0.8mm)}}}

\newcommand{\rectangleblue}{\raisebox{0pt}{\tikz{\draw[blue,solid,line width = 1.0pt](2.mm,0) rectangle (3.5mm,1.5mm);\draw[-,blue,solid,line width = 1.0pt](0.,0.8mm) -- (5.5mm,0.8mm)}}}

\title{Asymptotic Performance Analysis of NOMA Uplink Networks Under Statistical QoS Delay Constraints}
\author{\IEEEauthorblockN{Mouktar Bello\IEEEauthorrefmark{1}$,$ Arsenia Chorti\IEEEauthorrefmark{1}, Inbar Fijalkow\IEEEauthorrefmark{1}, Wenjuan Yu \IEEEauthorrefmark{2} and Leila Musavian\IEEEauthorrefmark{4}} \\
\IEEEauthorblockA{\IEEEauthorrefmark{1}ETIS UMR8051, CY Cergy Paris University, ENSEA, CNRS, F-95000, Cergy, France}

\IEEEauthorblockA{\IEEEauthorrefmark{2}5GIC, Institute of Communication Systems, University of Surrey, Guildford, GU2 7XH, UK}

\IEEEauthorblockA{\IEEEauthorrefmark{4}School of Computer Science and Electronic Engineering, University of Essex, Colchester, CO4 3SQ, UK}
}




\maketitle

\begin{abstract}
In this paper, we study the performance of an uplink non-orthogonal multiple access (NOMA) network under statistical quality of service (QoS) delay constraints, captured through each user's effective capacity (EC). We first propose novel closed-form expressions for the EC in a two-user NOMA network and show that in the high signal-to-noise ratio (SNR) region, the ``strong'' NOMA user, referred to as $U_2$, has a limited EC, assuming the same delay constraint as the ``weak'' user, referred to as $U_1$.  We demonstrate that for the weak user $U_1$, OMA and NOMA have comparable performance at low transmit SNRs,  while  NOMA  outperforms  OMA  in  terms  of  EC  at  high SNRs.  On the other hand, for the strong user $U_2$, NOMA achieves higher EC than OMA at small SNRs, while OMA becomes more beneficial at high SNRs. Furthermore, we show that at high transmit SNRs, irrespective of whether the application is delay tolerant, or not, the performance gains of NOMA over OMA for $U_1$, and OMA over NOMA for $U_2$ remain 
unchanged. When the delay QoS of one user is fixed, the performance gap between NOMA and OMA in terms of total EC increases with decreasing statistical delay QoS constraints for the other user. Next, by introducing pairing, we show that NOMA with user-pairing outperforms OMA, in terms of total uplink EC. The best pairing strategies are given in the cases of four and six users NOMA, raising once again the importance of power allocation in the optimization of NOMA's performance.
\end{abstract}


\section{Introduction}
 
\textcolor{black}{Non-orthogonal} multiple access (NOMA) schemes have attracted a lot of attention recently, allowing multiple  users to be served simultaneously with enhanced spectral efficiency; it is known that the boundary of achievable rate pairs using NOMA is outside the capacity region achievable with orthogonal multiple access (OMA) techniques \cite{islam2016power,makki2020survey,7973146,6666209,6868214}. Superior achievable rate are attainable \textcolor{black}{through} the use of superposition coding at the transmitter and of successive interference cancellation  (SIC) at the receiver \cite{saito2013non}. The SIC receiver decodes multi-user signals with descending received signal power and subtracts the decoded signal(s) from the received \textcolor{black}{superimposed} signal, so as to improve the signal-to-interference ratio. The process is repeated until the signal of interest is decoded \cite{higuchi2015non}. The interest in NOMA is linked to the multiple possibilities it offers, for example, in massive machine type communications (mMTC) systems where a large number of smart internet of things (IoT) devices try to access the shared resources simultaneously.

In uplink NOMA networks, the strongest user's signal is decoded first (reverse order with respect to the downlink). However the use of SIC limits the promised performance gain brought by NOMA due to the error propagation \cite{8438862,6182560,8011071}. Authors in \cite{8918153} introduced an iterative interference cancellation (IIC) detection scheme for uplink NOMA, and proposed a new detection scheme based on IIC, which is called advanced IIC (AIIC). Its shown that the bit error rate performance of AIIC is much better than that of SIC.

Similarly, the combination of NOMA with other emerging techniques and technologies such as new modulation techniques, user pairing, resource allocation algorithms (power and channel), MIMO, etc., improves its performance \cite{8085125,8010756,8352621,7095538,7974731}. Furthermore, NOMA offers a natural scenario for physical layer security as one user's signal is naturally degraded with respect to the other's \cite{Wenjuan-Chorti} and constitutes the equivalent of a helping interferer \cite{Chorti-Poor}.

Besides, in a number of emerging applications, delay constraints become increasingly important, e.g., ultra reliable low latency communication (URLLC) systems such as autonomous vehicles and enhanced reality.  Furthermore, in future wireless networks, users are expected to necessitate flexible delay guarantees for achieving different service requirements. In order to satisfy diverse delay requirements, a simple and flexible delay quality of service (QoS) model is imperative to be applied and investigated. In this respect, the effective capacity (EC) theory can be employed \cite{yu2016tradeoff,wu2003effective, tang2007cross}. The EC denotes the average maximum constant arrival rate which can be served by a given service process, while guaranteeing the required statistical delay provisioning \cite{musavian2015effective}. 

The delay-constrained communications for a downlink NOMA network was studied in \cite{yu2018link}, where the EC theory was utilized. The present analysis on uplink complements \cite{yu2018link} which focused on downlink transmissions. NOMA, as a more spectrum-efficient technique, is considered to be promising for supporting the massive number of devices to access the uplink connections. 

 In this paper, we provide a performance evaluation of the uplink transmission for a two-user NOMA network and a NOMA network with multiple user pairs under delay constraints, captured through the users' \textcolor{black}{ECs}. We note that the EC is a QoS aware \textcolor{black}{data-link} layer metric \cite{wu2003effective}, that captures the achievable rate under a delay violation probability threshold. We first derive novel closed-form expressions for a two-user network; we then provide asymptotic analysis for the network with NOMA and OMA.
 The conclusions drawn are supported by an extensive set of simulations. 

The paper is organized as follows. In Section II, we investigate the notion of EC in an uplink NOMA system under delay QoS constraints. In Section III, an asymptotic analysis on that metric is provided in a two-user system. In Section IV, the EC of multiple pairs is presented to investigate the impact of pairing. Simulation results are given in Section V, followed by conclusions in Section VI.

\section{Effective Capacity in Uplink NOMA}
\subsection{General Case: $M$-User NOMA}
Assume a $M$-user NOMA uplink network with users $U_1,U_2,...U_M$ in Rayleigh block-fading propagation channels \cite{923715}, with respective channel gains during a transmission block denoted by  $|h_i|^2,i=1,\dots,M$, that without loss of generality are ordered as $|h_1|^2<\dots<|h_M|^2$. The users transmit corresponding unit power symbols $s_1,\dots,s_M$ respectively, with
$\mathbb{E}[|s_i|^2]=1,i=1,\dots,M$ with a total transmit power constraint $P_T=\sum_{i=1}^{M}{P_i}=1$. We note in passing that the total power constraint does not capture the individual user's budgets, but rather regulatory requirements imposing that the transmit power in any given resource block cannot exceed a maximum value \cite{yang2016uplinknoma}. The received superimposed signal can be expressed as\cite{nzhang2016uplink}:

\begin{equation}
    z = \sum_{i=1}^{M}\sqrt{P_i}h_is_i+w,
\end{equation}
where \textcolor{black}{$w$} denotes a zero mean circularly symmetric complex Gaussian random variable with variance $\sigma^2$, i.e., $w \sim CN(0, \sigma^2)$.  
The receiver first decodes the symbols of the strongest user treating the transmission of the weaker users as interference. After decoding it, the receiver suppresses it from $z$ and decodes the signal of the second strongest user, and so on until the decoding of the weakest user's signal. Following the SIC principle and denoting the transmit SNR $\rho = \frac{1}{\sigma^2}$, the achievable rate, in b/s/Hz, for user
$U_i, i=1,\dots,M$,  assuming no error propagation, is expressed as \cite{fan2015uplink}:
 \begin{equation}
    R_i= \log_2 \left(1+ \frac{\rho P_i|h_i|^2}{1+\rho \sum_{l=1}^{i-1}P_l|h_l|^2}\right).
 \end{equation}
 
Next, let $\theta_i$ \textcolor{black}{be} the statistical delay exponent of the $i$-th user, i.e., $\theta_i$ captures how strict the delay constraint of the user $i$ is, and assume that the service process satisfies the G\"{a}rtner-Ellis theorem \cite{wu2003effective}. A slower decay rate can be represented by a smaller $\theta_i$, which indicates that the system is more delay tolerant, while a larger $\theta_i$ corresponds to a system with more stringent QoS requirements. 
Applying the EC theory in an uplink NOMA with $M$ users, the $i$-th user's EC over a block-fading channel is defined as:
\begin{align}
      E_c^i= -\frac{1}{\theta_i T_{{f}} B} \ln \bigg( \mathbb{E}\left[e^{-\theta_i T_{\text{f}} B R_i}\right] \bigg) \quad \left(\text{in b/s/Hz}\right),
      \label{eq:ECdefinition}
\end{align}where $T_f$ is the block duration, $B$ is the bandwidth and $\mathbb{E}\left[\cdot\right]$ denotes expectation over the channel gains. 
By inserting $R_i$ into (\ref{eq:ECdefinition}), we obtain the following expression for the EC of the $i$-{th} user
\begin{align}
&E_c^i = \frac{1}{\beta_i}\log_2\left(\mathbb{E}\left[\left(1+ \frac{\rho P_i|h_i|^2}{1+\rho \sum_{l=1}^{i-1}P_l|h_l|^2}\right)^{\beta_i}\right]\right)
\label{eq:Eci}
\end{align}
where $\beta_i=-\frac{\theta_iT_fB}{\ln{2}}$, $ i=1,\dots M$, is the normalized (negative) QoS exponent. Developing (\ref{eq:Eci}), we have that:
\begin{align}
&E_c^i=\frac{1}{\beta_i}\log_2\Bigg(\int_{0}^{\infty} \int_{x_1}^{\infty} \int_{x_2}^{\infty}...\int_{x_{i-1}}^{\infty}\!\!\left(1+\frac{\rho P_{i}x_i}{1+\sum_{l=1}^{i-1}\rho P_{l}x_l}\right)^{\beta_i}\nonumber\\ &f_{_{X(1)},_{X(2)},\dots,_{X(i)}}\left(x_1,x_2,...,x_i\right)d_{x_i}\ d_{x_i-1}...d_{x_1} \Bigg),
\label{Ecidev}
\end{align}where $f_{_{X(1)},_{X(2)},\dots,_{X(i)}}\left(x_1,x_2,...,x_i\right)$ is the joint distribution of $x_i=|h_i|^2, i=1,\dots,M$.

To evaluate the joint distribution of the channel gains, we make use of the theory of order statistics in the following analysis \cite{yang2011order}.
The probability density function (PDF) of the $i$-th ordered random variable in a population of $M$ is given by:
\begin{align}
    f_{_{X(i)}}(x) = \psi_i f(x) (1-F(x))^{M-i}F(x)^{i-1},
    \label{pdf}
\end{align}
where \(\psi_i=\frac{1}{B(i,M-i+1)}\), and, $B(a,b)$ is the beta function $B(a,b)=\frac{\Gamma(a)\Gamma(b)}{\Gamma(a+b)}$, with  $\Gamma(a)=(a-1)!$. Assuming a Rayleigh wireless environment, the channel gains, denoted by $x_i=|h_i|^2$, are exponentially distributed with PDF and cumulative density function (CDF) respectively given by $f(x)=e^{-x},$ and $F(x)=1-e^{-x}.$ 

The joint distribution of $M$ order statistics is given by \cite{yang2011order}:
\begin{align}
    f_{_{X(1)\dots X(M)}}(x_1,x_2,\dots,x_M)=M!f_{_{X(1)}}(x_1)\dots f_{_{X(M)}}(x_M),
\end{align}
where $\textbf{  } x_1\leq x_2 \leq\dots\leq x_M$, while for any two order statistics, we have that:
\begin{align}
&f_{_{X(l)},_{X(k)}}(x_l,x_k)=\frac{M!}{(l-1)!(k-l-1)!(M-k)!}\nonumber\\
&\times(1-F(x))^{l-1}f(x)(F(x)-F(y))^{k-l-1}f(y)(F(y))^{M-k}.
\label{pdfjoint}
\end{align}

\subsection{Case of Two-User NOMA Uplink Network ($M$=2)}
Using \eqref{pdf}, we obtain 
\begin{align}
f_{_{X(1)}}(x_1)&=2e^{-2 x_1}.
\label{eq:joint pdf 2 user}
\end{align}
Furthermore, by setting $M=2, l=1$ and $k=2$ in \eqref{pdfjoint}, we get:
\begin{align}
   &f_{_{X(1)},_{X(2)}}(x_1, x_{2}) = 2 f(x_{1})f(x_{2}) = 2 e^{-x_{1}}e^{-x_{2}} .
  \label{eq12}
\end{align}
As a result, the EC of $U_1$, denoted by $E_c^1$, is expressed as
\begin{eqnarray}
    E_c^1 &=&\frac{1}{\beta_1}\log_2\bigg(\mathbb{E}[(1+ \rho P_1 x_{1})^{\beta_1}]\bigg) \nonumber
    \\ 
    &=& \frac{1}{\beta_1}\log_2\left(\int_{0}^{\infty}\left(1+ \rho P_1 x_1\right)^{\beta_1}f_{_{X(1)}}(x_1) dx_1\right) \nonumber\\
    &=&\frac{1}{\beta_1}\log_2\left(\frac{2}{P_1 \rho}\times U\left(1,2+\beta_1,\frac{2}{\rho P_1}\right)\right) .
    \label{eq:EC1}
\end{eqnarray}
where $U(\cdot, \cdot, \cdot)$ denotes the confluent hypergeometric function \cite{yu2018link}.\newline
On the other hand, the EC of $U_2$ is evaluated as 
\begin{align}
  &E_c^2 = \frac{1}{\beta_2}\log_2\left(\mathbb{E}\left[\left(1+ \frac{\rho P_2 x_{2}}{1+ \rho P_1 x_{1}}\right)^{\beta_2}\right]\right)\nonumber \\
  &=\!\!\frac{1}{\beta_2}\!\!\log_2\left(\!\!\int_{0}^{\infty}\!\!\!\!\int_{x_{1}}^{\infty}\!\!\!\!\Big(1+\!\! \frac{\rho P_2 x_2}{1+\rho P_1 x_1}\Big)^{\beta_2}\!f_{_{X(1)},_{X(2)}}(x_1,x_2) dx_2 dx_1\!\!\right) \nonumber\\
  &=\frac{1}{\beta_2}\log_2\bigg(2 P_2^{1-\beta_2}(\rho P_2)^{\beta_2} e^{\frac{1}{\rho P_2}}e^{-\frac{(P_1-P_2)}{\rho P_2}}\bigg)
  \nonumber \\
  &+\frac{1}{\beta_2}\log_2\Bigg(\sum_{j=0}^{-\beta_2}\binom{-\beta_2}{j}(\rho P_1)^{j}\times  \sum_{k=0}^{\infty}\frac{(-1)^k (P_2-P_1)^k}{k!(1+j+k)}\nonumber \\
  &\!\!\times\!\!\Big[\Gamma\left(\!\!2+\beta_2+j+k,\frac{1}{\rho P_2}\!\!\right)\!\!
  -\!(\rho P_2)^{-1-j-k}\Gamma\left(1+\beta_2,\frac{1}{\rho P_2}\right)\!\!\Big]\!\!\Bigg)
  \label{eq:EC2}
\end{align}
with $\Gamma(\cdot, \cdot)$ denoting the incomplete Gamma function \cite{yu2018link}.

\begin{IEEEproof}  
The proof is provided in Appendix I. 
\end{IEEEproof}

\subsection{Case of a Two-User OMA Network}
Similarly, using time division multiple access (TDMA), the achievable data rate of the $i$-th user in a two-user OMA network, denoted by $\widetilde{R}_{i}, i=1,2$, is given by
\begin{equation}
  \widetilde{R}_{i}=\frac{1}{2}\log_2\bigg(1+2\rho P_i|h_i|^2\bigg), i=1,2.
\end{equation}
Note that $\frac{1}{2}$ is due to the equal allocation of resources to both users. Furthermore, it is important to note that the power of each OMA user is double that of NOMA, for the sake of fairness \cite{yu2018link}. The corresponding ECs of both users in an OMA network are denoted by $\widetilde{E}_c^{i}$:
\begin{align}
&\widetilde{E}_c^{i}=\frac{1}{\beta_i}\log_2\bigg(\mathbb{E}\Big[(1+2\rho P_i  |h_i|^2)^{\frac{\beta_i}{2}}\Big] \bigg)\label{eq:EC1OMA}.
\end{align}
A general expression of the ECs of $M$ TDMA OMA users is given in \cite{yu2018link}; applying this to a two-user network we can be easily obtain:
\begin{align}
&\widetilde{E}_c^{1}=\frac{1}{\beta_1}\log_2\left(\frac{1}{\rho P_1}\times U\left(1, 2+\frac{\beta_1}{2}, \frac{1}{\rho P_1}  \right)\right),\\
&\widetilde{E}_c^{2}=\!\frac{1}{\beta_2}\log_2\!\left(\!\frac{1}{\rho P_2}\!\sum_{k=0}^{1}\binom{1}{k}(-1)^k \times U\!\left(1, 2+\frac{\beta_2}{2}, \frac{1+k}{2\rho P_2}  \right)\!\right).
 \end{align}
The difference in these expressions is due to the different PDFs of ordered channel gains.
\section{Asymptotic Analysis}
In this Section, an asymptotic analysis with respect to the transmit SNR $\rho$ is presented. This analysis consists in describing the limiting behavior of individual and total ECs, and how they evolve with the transmit SNR $\rho$. Our results are summarized in the following Propositions and Lemmas. 
\subsection{Case 1: Delay-Constrained Users}
\begin{prop}
\begin{enumerate}
\item At low transmit SNR, $\rho \rightarrow0$,  $E_c^1$, $\widetilde{E}_c^{1}$, $E_c^2$ and $\widetilde{E}_c^{2}$   start at zero and then increase at the same rate for any user.
\item At high values of the transmit SNR, $\rho>>1$, $E_c^1$ increases faster than $\widetilde{E}_c^{1}$ and NOMA becomes more advantageous than OMA, for $U_1$. While for $U_2$, $\widetilde{E}_c^{2}$  increases faster than $E_c^2$, although NOMA is outperforming OMA.
\item At very high values of the transmit SNR, $\rho\rightarrow\infty$, the performance gain of NOMA over OMA increases at gradually reducing rate, for $U_1$. Albeit, for $U_2$, $E_c^2$  reaches an upper limit, allowing OMA to outperform NOMA after some SNR value (which depends on the system parameters).
\end{enumerate}
\end{prop}
\textbf{\textit{Proposition 1}} is the synthesis of \textbf{\textit{Lemmas 1, 2 and 3}}, discussed in detail next.
\begin{lemma}\textit{In the low and high SNR regimes, respectively, the following conclusions hold:
        \begin{enumerate}
            \item When $\rho\rightarrow0$, then, $E_c^1\rightarrow0$, $E_c^2\rightarrow0$, $\widetilde{E}_c^{1}\rightarrow0$, $\widetilde{E}_c^{2}\rightarrow0$, $E_c^1-\widetilde{E}_c^{1}\rightarrow0$, $E_c^2-\widetilde{E}_c^{2}\rightarrow0$;
            \item When $\rho\rightarrow +\infty$, then $E_c^1\rightarrow+\infty$, $E_c^2\rightarrow \frac{1}{\beta_2}\log_2\left(\mathbb{E}\left[\left(1+\frac{P_2 |h_2|^2}{P_1 |h_1|^2}\right)^{\beta_2}\right]\right)$, $\widetilde{E}_c^{1}\rightarrow+\infty$, $\widetilde{E}_c^{2}\rightarrow+\infty$, $E_c^1-\widetilde{E}_c^{1}\rightarrow+\infty$, $E_c^2-\widetilde{E}_c^{2}\rightarrow-\infty$.
        \end{enumerate}}
\end{lemma}

\begin{IEEEproof} The proof is provided in Appendix II.
\end{IEEEproof}

To further analyze the impact of $\rho$ on the individual EC, the partial derivatives with the respect of $\rho$ are investigated \cite{yu2018link}.
\begin{lemma} \textit{For the EC of the $U_{1}$, in a two-user uplink network the following hold: 
             \begin{enumerate}
              \item $ \frac{\partial E_c^1}{\partial\rho}\ge0$ and $\frac{\partial\widetilde{E}_c^{1}}{\partial\rho}\ge0$, $\forall \rho$;
              \item When $\rho\rightarrow0$, then $\lim\limits_{\rho\rightarrow0}(\frac{\partial (E_c^1-\widetilde{E}_c^{1})}{\partial\rho})=0$;
              \item When $\rho>> 1$, then $\frac{\partial (E_c^1-\widetilde{E}_c^{1})}{\partial\rho}\approx\frac{1}{2\rho \ln2}\ge0$ and it approaches $0$ when $\rho\rightarrow\infty$.
             \end{enumerate}
}
        \end{lemma}
       \begin{IEEEproof} The proof is provided in Appendix III.\end{IEEEproof}


\begin{lemma} \textit{For the EC of the $U_{2}$, in a two-user uplink network the following hold: 
             \begin{enumerate}
              \item $ \frac{\partial E_c^2}{\partial\rho}\ge0$ and $ \frac{\partial \widetilde{E}_c^{2}}{\partial\rho}\ge0$, $\forall \rho$;
              \item When $\rho\rightarrow0$, then $\lim\limits_{\rho\rightarrow0}(\frac{\partial (E_c^2-\widetilde{E}_c^{2})}{\partial\rho})=0$
              \item When $\rho>> 1$, then $\frac{\partial (E_c^2-\widetilde{E}_c^{2})}{\partial\rho}\approx-\frac{1}{2 \ln2}\frac{1}{\rho}<0$ and it approaches 0 when $\rho\rightarrow\infty$.
             \end{enumerate}
      }
        \end{lemma}
      \begin{IEEEproof}  The proof is provided in Appendix IV.\end{IEEEproof}
     

    
Finally, we investigate the sum ECs when using OMA and NOMA, denoted by $V_N$ and $V_O$, respectively, i.e.,
    \begin{eqnarray}
        V_N&=&E_c^1+E_c^2,\\
        V_O&=&\widetilde{E}_c^1+\widetilde{E}_c^2.
    \end{eqnarray}
\begin{prop}
\begin{enumerate}
\item At low transmit SNR $\rho$, $V_N$ and $V_O$ increase at a constant rate that depends on the average of the channel power gains and the allocated power coefficients.
\item When $\rho>>1$, $V_N$ and $V_O$ tend to $\infty$, and reach a plateau when the transmit SNR $\rho\rightarrow\infty$.
\end{enumerate}
\end{prop}
    \textbf{\textit{Proposition 2}} is the consequence of the \textbf{\textit{Lemma 4}}.
\begin{lemma} \textit{For the sum EC with NOMA, denoted by $V_N$, and with OMA, denoted by $V_O$, in a two-user uplink network, the following hold:
              \begin{enumerate}
              \item $ \frac{\partial V_N}{\partial\rho}\ge0$ and $\frac{\partial V_O}{\partial\rho}\ge0$, $\forall \rho$;
              \item When $\rho\rightarrow0$, $V_N\rightarrow0$, $ \lim\limits_{\rho\rightarrow 0}(\frac{\partial V_N}{\partial\rho})=\frac{P_1}{\ln2}\mathbb{E}[|h_1|^2]+\frac{P_2}{ \ln2}\mathbb{E}[|h_2|^2]\ge0$, and $V_O\rightarrow0$, $ \lim\limits_{\rho\rightarrow 0}(\frac{\partial V_O}{\partial\rho})=\frac{P_1}{\ln2}\mathbb{E}[|h_1|^2]+\frac{P_2}{ \ln2}\mathbb{E}[|h_2|^2]\ge0$;
              \item When $\rho>>1$, $V_N\rightarrow\infty$, $\lim\limits_{\rho\rightarrow\infty}(\frac{\partial V_N}{\partial\rho})=0$, and $V_O\rightarrow\infty$, $\lim\limits_{\rho\rightarrow\infty}(\frac{\partial V_O}{\partial\rho})=0$.
             \end{enumerate}
}
\end{lemma}
\begin{IEEEproof} The proof is provided in Appendix V.\end{IEEEproof}

\subsection{Case 2: Delay-Tolerant Applications}
A case of particular interest is presented when the users' applications are delay tolerant, i.e., when the delay exponent becomes negligible. 
In this case, investigation of the ECs of the two-user, uplink NOMA and OMA networks, is performed without delay constraints.  The impact of the transmit SNR $\rho$ in this case is also investigated.

\begin{prop}
\begin{enumerate}
\item For both OMA and NOMA, when there is no delay constraint ($\theta=0$), the individual ECs of both users are equal to their ergodic capacities.
\item At high transmit SNRs, irrespective of whether there's a tolerance for delay or not, the conclusions on the performance gain of NOMA over OMA for $U_1$, and OMA over NOMA for $U_2$ remain the same.
\end{enumerate}
\textbf{\textit{Proposition 3}} is the consequence of the \textbf{\textit{Lemma 5}}.
\end{prop}
\begin{lemma} \textit{Considering the EC for the weaker user with $\theta_1\rightarrow0$, in NOMA and OMA, the following hold:}
             \begin{enumerate}
              \item[a)] When $\theta_1\rightarrow0$, $\lim\limits_{\theta_1\rightarrow0}E_c^1=\mathbb{E}[R_1],$ $\lim\limits_{\theta_1\rightarrow0}\widetilde{E}_c^{1}=\mathbb{E}[\widetilde{R}_1],$ $\lim\limits_{\theta_1\rightarrow0}(E_c^1-\widetilde{E}_c^{1})=\mathbb{E}[R_1]-\mathbb{E}[\widetilde{R}_{1}],$
              \item[b)] When $\theta_1\rightarrow0, \rho\rightarrow\infty$,       $\lim\limits_{\substack{\theta_1\rightarrow0\\\rho\rightarrow\infty}}E_c^1=\infty$, $\lim\limits_{\substack{\theta_1\rightarrow0\\\rho\rightarrow\infty}}\widetilde{E}_c^{1}=\infty$, $\lim\limits_{\substack{\theta_1\rightarrow0\\\rho\rightarrow\infty}}(E_c^1-\widetilde{E}_c^{1}) =\infty$.
            \end{enumerate}              
\textit{Considering the EC for the stronger user with $\theta_2\rightarrow0$, in NOMA and OMA, we prove that}:
             \begin{enumerate}
              \item[c)] When $\theta_2\rightarrow0$, $ \lim\limits_{\theta_2\rightarrow0}E_c^2=\mathbb{E}[R_2]$,
              $ \lim\limits_{\theta_2\rightarrow0}\widetilde{E}_c^{2}=\mathbb{E}[\widetilde{R}_2]$, $\lim\limits_{\theta_2\rightarrow0}(E_c^2-\widetilde{E}_c^{2})=\mathbb{E}[R_2]-\mathbb{E}[\widetilde{R}_{2}]$,
              \item[d)] When $\theta_2\rightarrow0, \rho\rightarrow\infty$, \newline
              $\lim\limits_{\substack{\theta_2\rightarrow0\\\rho\rightarrow\infty}}E_c^2=\mathbb{E}\bigg[\log_2\bigg(1+ \frac{P_2 |h_2|^2}{P_1 |h_1|^2}\bigg)\bigg]$, $\lim\limits_{\substack{\theta_2\rightarrow0\\\rho\rightarrow\infty}}\widetilde{E}_c^{2}=\infty$, $\lim\limits_{\substack{\theta_2\rightarrow0\\\rho\rightarrow\infty}}(E_c^2-\widetilde{E}_c^{2}) =-\infty$.
             \end{enumerate}
             

\end{lemma}

\begin{IEEEproof} The proof is provided in Appendix VI.\end{IEEEproof}


 \section{Effective Capacity of Multiple NOMA Pairs}
  The $M$ NOMA users scenario assumes that the resource block is shared among $M$ users. For large values of $M$, stronger users are penalized due to high interference level from weaker users since they are decoded first \cite{zhu2018optimal}. Pairing allows us to mitigate interference from weaker users on stronger ones. A popular approach for alleviating this effect in an $M$ user network, is to form $\frac{M}{2}$ groups with indices $i=1,\dots,\frac{M}{2}$, where each group involves only $2$ users.
  Inside each group, NOMA is implemented, while across different groups TDMA is applied.
  
 The achievable data rate of the two users, $U_1$ and $U_2$ of the $i^{th}$group, where $|h_{1_i}|^2\leq|h_{2_i}|^2$, can be formulated as follow:
   \begin{equation}
   R_{1_i}= \frac{2}{M}\log_2 \left(1+ \rho P_{1_i}|h_{1_i}|^2\right),
   \label{eq:r1i}
 \end{equation}
   \begin{equation}
    R_{2_i}=\frac{2}{M} \log_2 \left(1+ \frac{\rho P_{2_i}|h_{2_i}|^2}{1+\rho P_{1_i}|h_{1_i}|^2}\right),
    \label{eq:r2i}
\end{equation}
 with $\frac{2}{M}$ the fraction of resources at the disposal of the two users inside a NOMA group.
 
  On the other hand, if all users utilize TDMA, their achievable data rates are given as follows:
    \begin{equation}
   \tilde{R}_{j}= \frac{1}{M}\log_2 \left(1+ 2 P_j\rho |h_{j}|^2\right), j\in \left\{1_i,2_i\right\}.
   \label{eq:rj}
    \end{equation}
 The factor $\frac{1}{M}$ is to indicate that each user has only one time slot to transmit. 
 
By replacing (\ref{eq:r1i}) and (\ref{eq:r2i}) in (\ref{eq:ECdefinition}), we get respectively the following ECs for $U_1$ and $U_2$ in the $i^{th}$ group:
\begin{eqnarray}
   E_c^{1_i} &=& \frac{1}{\beta_{1_i}}\log_2\left(\mathbb{E}\bigg[(1+ \rho P_{1_i}|h_{1_i}|^2)^{\frac{2 \beta_{1_i}}{M}}\bigg]\right),
\end{eqnarray}

 \begin{equation}
   E_c^{2_i} = \frac{1}{\beta_{2_i}}\log_2\left(\mathbb{E}\bigg[\left(1+ \frac{\rho P_{2_i}|h_{2_i}|^2}{1+ \rho P_{1_i}|h_{1_i}|^2}\right)^{\frac{2 \beta_{2_i}}{M}}\bigg]\right).
\end{equation}

On the other hand, replacing (\ref{eq:rj})  in (\ref{eq:ECdefinition}) we get the expressions for both users while using TDMA:
\begin{eqnarray}
  \widetilde{E}_c^{1_i} &=& \frac{1}{\beta_{1_i}}\log_2\left(\mathbb{E}\bigg[(1+ 2\rho P_{1,i} |h_{1_i}|^2)^{\frac{\beta_{1_i}}{M}}\bigg]\right),
\end{eqnarray}

\begin{eqnarray}
 \widetilde{E}_c^{2_i} &=& \frac{1}{\beta_{2_i}}\log_2\left(\mathbb{E}\bigg[(1+2\rho P_{2,i} |h_{2_i}|^2)^{\frac{\beta_{2_i}}{M}}\bigg]\right).
\end{eqnarray}
Next, we analyze the total sum EC of multiple NOMA pairs, denoted by $E_c^{tot}$, in comparison with the total sum EC for the $M$ OMA users, $\tilde{E_c}^{tot}$ defined as:
\begin{align}
&E_c^{tot}=\sum_{i=1}^{\frac{M}{2}}(E_c^{1_i}+E_c^{2_i}),\\
&\tilde{E_c}^{tot}=\sum_{i=1}^{\frac{M}{2}}(\widetilde{E}_c^{1_i}+\widetilde{E}_c^{2_i}).
\end{align}

To investigate the performance of the user-pairing, the following Proposition and Lemma are provided.

\begin{prop}
\begin{enumerate}
\item NOMA user-pairing outperforms OMA  at low transmit  SNRs and this performance gain carries on at very high transmit SNRs, with the possibility to be improved by optimizing the power allocation.
\end{enumerate}
\end{prop}
\textbf{\textit{Proposition 4}} is the consequence of \textbf{\textit{Lemma 6}}.
\begin{lemma}Considering $E_c^{tot}-\tilde{E_c}^{tot}$, we prove that:
            \begin{enumerate}
             \item [a)]When $\rho\rightarrow0$, $E_c^{tot}-\tilde{E_c}^{tot}$$\rightarrow0$, and         $\lim\limits_{\rho\rightarrow0}\frac{\partial(E_c^{tot}-\tilde{E_c}^{tot})}{\partial\rho}=0.$ 
             
             \item [b)] When $\rho\rightarrow\infty$, $E_c^{tot}-\tilde{E_c}^{tot}$ $\rightarrow$ $constant$, given in \eqref{constant}, and
             $\lim\limits_{\rho\rightarrow\infty} \frac{\partial(E_c^{tot}-\tilde{E_c}^{tot})}{\partial\rho}=0.$ 
             
            \end{enumerate}
\end{lemma}
\begin{align}
&\lim\limits_{\rho\rightarrow\infty}(E_c^{tot}\!-\!\tilde{E_c}^{tot})\!\!=\!\!\sum_{i=1}^{\frac{M}{2}}\left(\!\frac{1}{\beta_{1,i}}\!\!\log_2\!\!\bigg(\!2^{-\frac{\beta_{1,i}}{M}} \mathbb{E}\Big[(  P_{1,i}|h_{1,i}|^2)^{\frac{\beta_{1,i}}{M}}\Big]\bigg) \right.\nonumber \\
&+\left.\frac{1}{\beta_{2,i}}\log_2\left(\frac{\mathbb{E}\Big[\left(1+ \frac{P_{2,i}|h_{2,i}|^2}{ P_{1,i}|h_{1,i}|^2}\right)^{\frac{2 \beta_{2,i}}{M}}\Big]}{\mathbb{E}\Big[( 2 P_{2,i} |h_{2,i}|^2)^{\frac{\beta_{2,i}}{M}}\Big]}\right)\right).
\label{constant}
\end{align}
\begin{IEEEproof} The proof is provided in Appendix VII.\end{IEEEproof}

From \textbf{\textit{Lemma 6}}, we can conclude that $E_c^{tot}-\tilde{E_c}^{tot}$ initially starts at 0, first increases at low transmit SNRs $\rho$, and finally approaches a $constant$ in \eqref{constant} that depends on the power allocation, at high transmit SNRs, i.e., this performance gain of NOMA with user-pairing over OMA can be optimized by finding the best pairing strategy.

\section{Numerical Results}
In this Section, the Propositions and Lemmas presented in previous sections are validated through Monte Carlo simulations.
We first consider a two-user uplink NOMA system, with the following parameters: normalized transmission power for both users, $P_1=0.2$, $P_2=0.8$, normalized delay exponent $\beta_1=\beta_2=-1$ for both users, unless otherwise stated. Fixed power allocation is used for the sake of simplicity, as the power control problem to maximize the sum effective capacity with delay QoS constraints is not treated in this contribution, while a suboptimal solution is proposed in \cite{choi2017effective}.

\begin{figure}[ht]
 \begin{center}
     \includegraphics[width=0.5\textwidth]{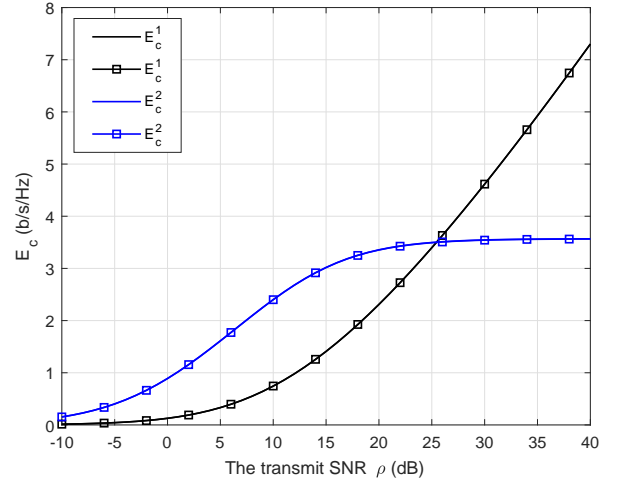}
     \caption{Validation of the closed-form expressions in uplink two-user NOMA system.}
      \label{fig5}
\end{center}
\end{figure}

 Fig.\ref{fig5} shows $E_c^1$ and $E_c^2$, with the closed-form expressions \rectangleblack, \rectangleblue, and the results of the Monte Carlo simulations \blackline , \blueline. So, the accuracy of the closed-form expressions is confirmed.
 
\begin{figure}[t]
  \begin{center}
\includegraphics[width=0.5\textwidth]{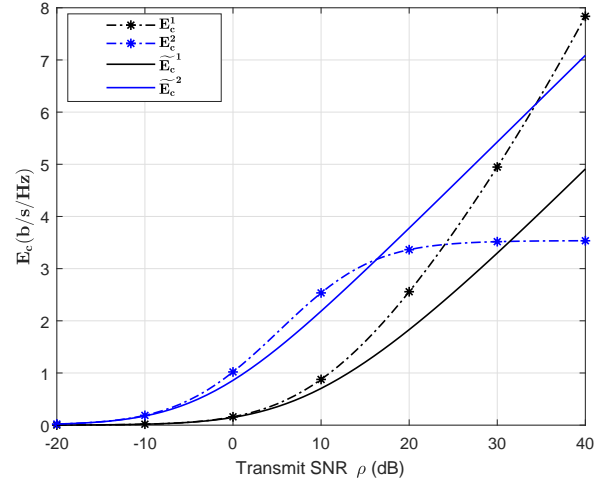}
\caption{
$E_c^1$, $E_c^2$, and $\tilde{E_c}^1$, $\tilde{E_c}^2$, versus the transmit SNR $\rho$}
      \label{fig6}
    \end{center}
\end{figure}

In Fig.\ref{fig6}, the ECs of the  two-user uplink NOMA and OMA networks are depicted  versus the transmit SNR. We note that for $U_1$, NOMA and OMA perform equally well at very low transmit SNRs, and NOMA is advantageous compared to OMA at high transmit SNRs. In contrast, for $U_2$, NOMA is better at low SNRs and OMA is advantageous at high transmit SNRs. We notice also that the EC of $U_2$ reaches a plateau at high SNRs, validating \textbf{\textit{Lemma 1}}. 
 
\begin{figure}[t]
     \begin{center}
      \includegraphics[width=0.5\textwidth]{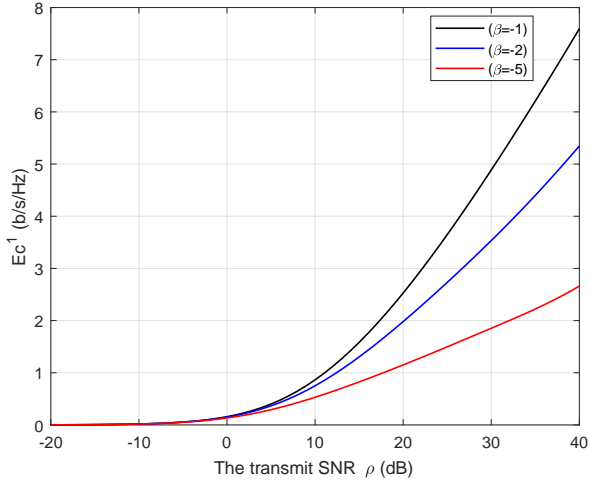}
      \caption{$E_c^1$ versus the transmit SNR, for different delay requirements.}
      \label{fig7}
       \end{center}
  \end{figure}

\begin{figure}[t]
       \begin{center}
      \includegraphics[width=0.5\textwidth]{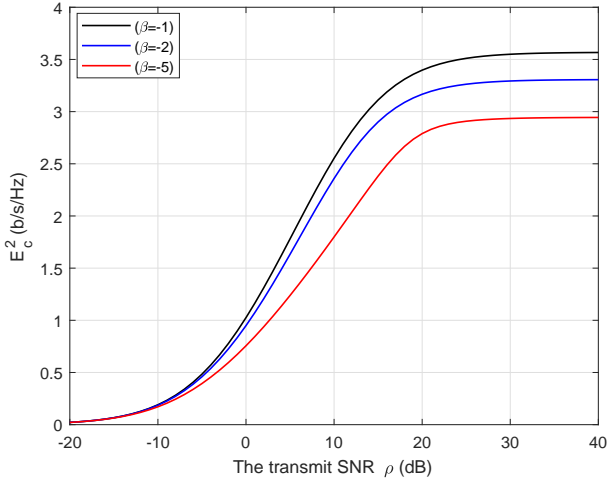}
      \caption{$E_c^2$ versus the transmit SNR $\rho$ for different delay requirements.}
      \label{fig8}
       \end{center}
  \end{figure}

 Fig.\ref{fig7} and Fig.\ref{fig8} show, respectively, the EC of $U_1$ and $U_2$, versus the transmit SNR, for different values of $\beta_1=\beta_2=\beta$. When the delay constraints become more stringent, i.e., $\beta$ decreases (equivalently, $\theta$ increases), the individual ECs in NOMA decrease, for both users. 

\begin{figure}[t]
  \begin{center}
     \includegraphics[width=0.5 \textwidth]{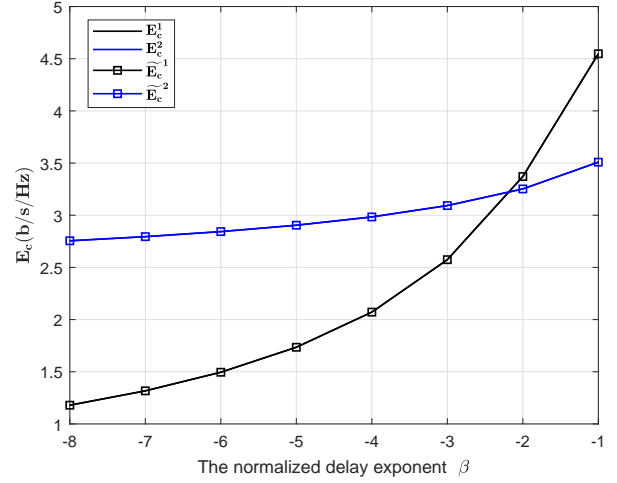}
     \caption{$E_c^1$, $E_c^2$, and $\tilde{E_c}^1$ and $\tilde{E_c}^2$, versus the (negative) normalized delay exponent $\beta$ at $\rho=30$ dB.}
      \label{figECbeta}
     \end{center}
 \end{figure}

\begin{figure}[t]
     \begin{center}
      \includegraphics[width=0.5 \textwidth]{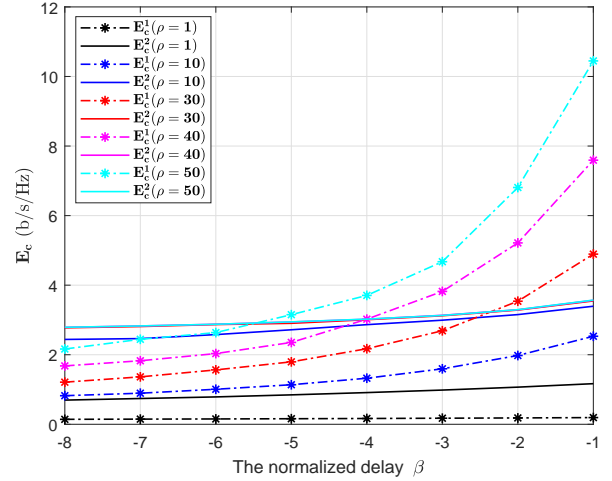}
      \caption{$E_c^1$, $E_c^2$ versus normalized delay $\beta$, for different values of $\rho$.}
      \label{fig10}
     \end{center}
  \end{figure}

In Fig. \ref{figECbeta}, the ECs of the strong and weak users are depicted in the high SNR regime ($\rho=30$ dB) as functions of the (negative) normalized delay exponent, for NOMA and OMA. We noticed that the EC curves are identical. On the other hand, in Fig.\ref{fig10} where $E_c^1$ and $E_c^2$ are depicted across different SNR values, ($\rho\in\{1, 10, 30, 40, 50\}$ dB, as functions of the (negative) normalized delay exponent, the EC of both users increase with the transmit SNR $\rho$ increasing.

\begin{figure}[t]
       \begin{center}
      \includegraphics[width=0.5\textwidth]{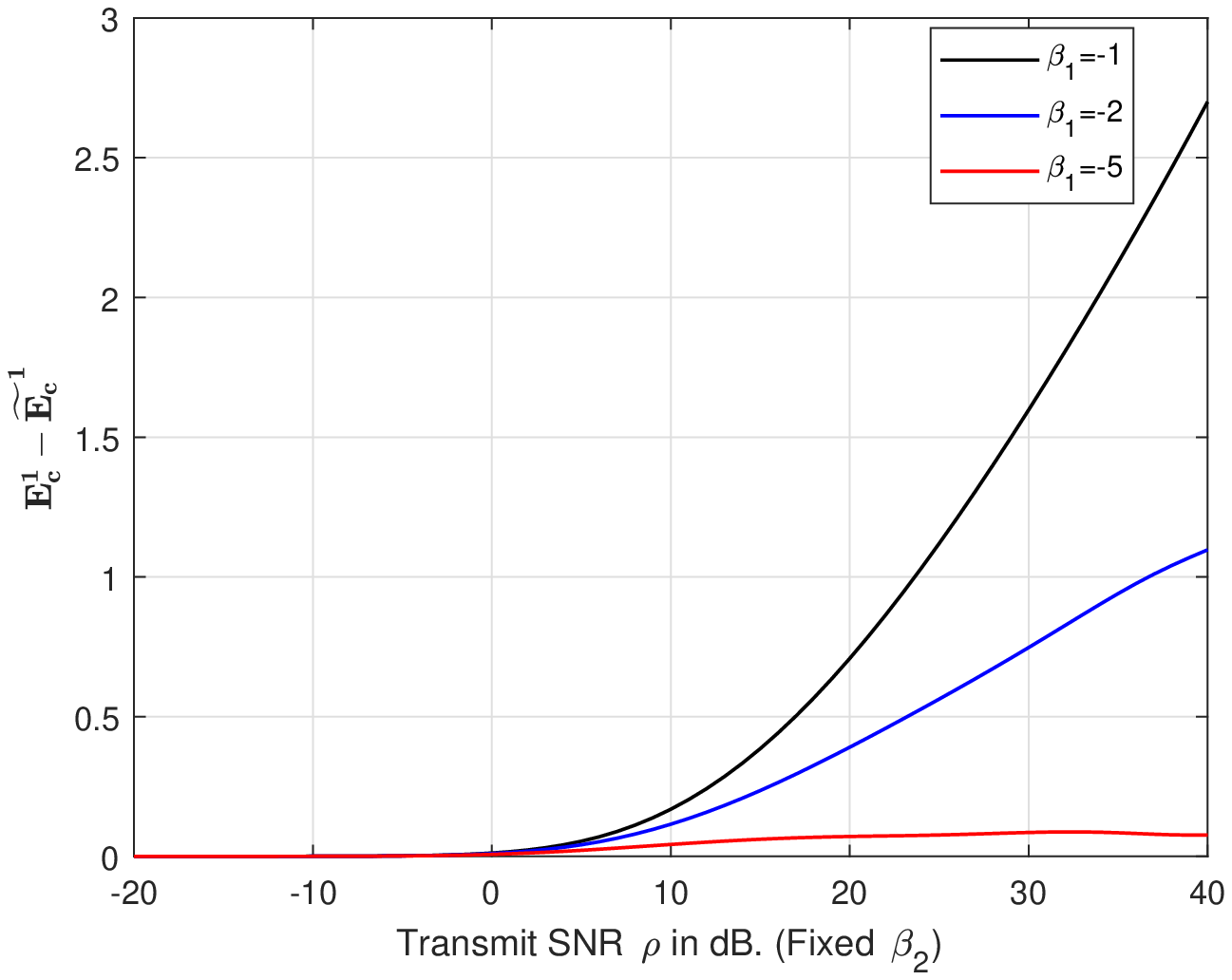}
      \caption{ $E_c^1-\Tilde{E}_c^1$ versus $\rho$, for several values of the normalized delay exponent.}
      \label{fig11}
     \end{center}
  \end{figure}

Fig. \ref{fig11} shows $E_c^1-\Tilde{E}_c^1$ versus the transmit SNR. This curve initially starts at zero, increases at the high transmit SNRs. Also, we can note that this gap decreases with delay constraints becoming more stringent ($\beta$ decreasing). This confirms \textbf{\textit{Lemma 2}}.
 
\begin{figure}[t]
  \begin{center}
      \includegraphics[width=0.5\textwidth]{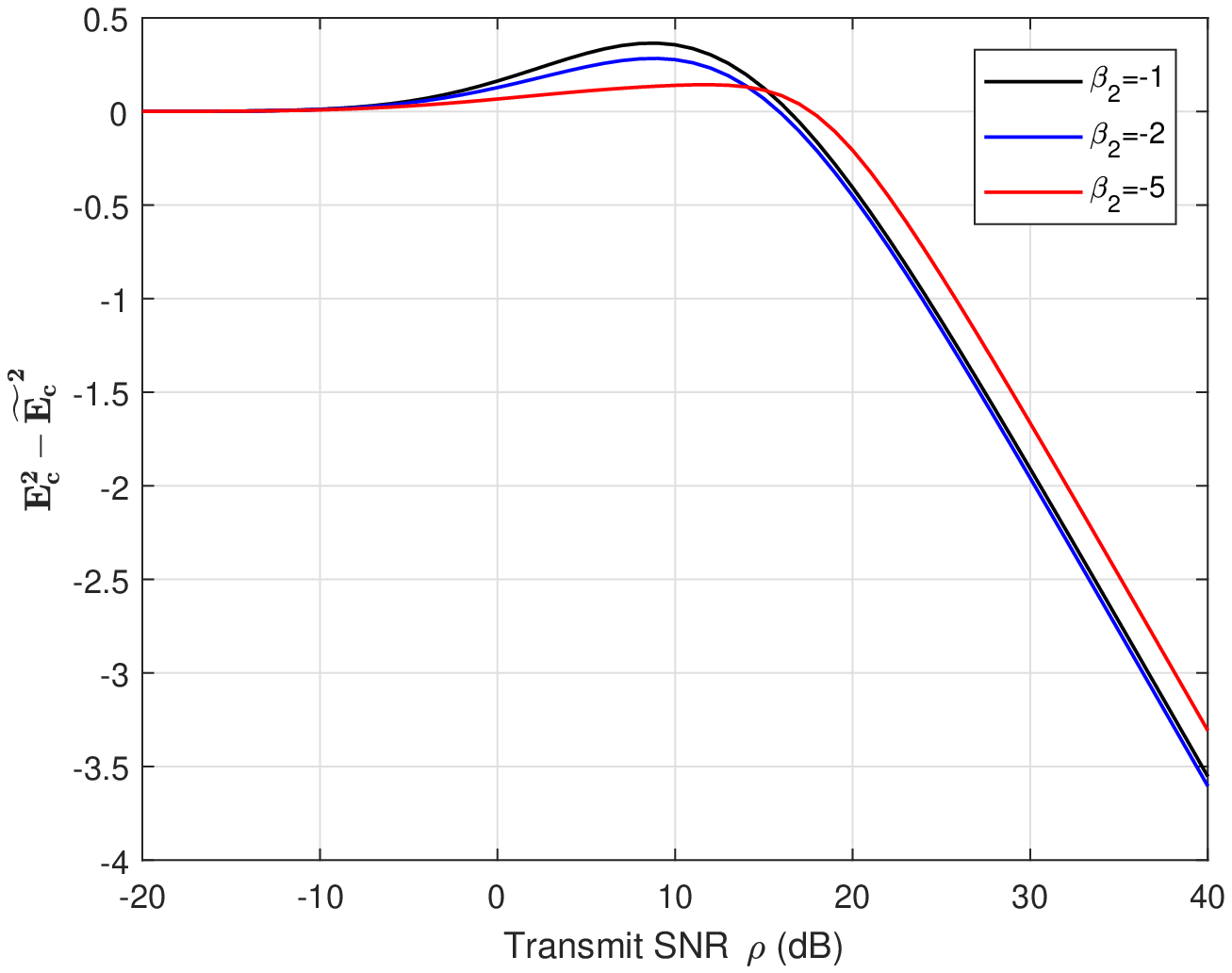}
      \caption{ $E_c^2-\Tilde{E}_c^2$ versus $\rho$, for various values of normalized delay exponent.}
      \label{fig12}
   \end{center}
  \end{figure}

Fig. \ref{fig12} shows $E_c^2-\Tilde{E}_c^2$ versus the transmit SNR. This curve initially starts at zero, increases to a certain maximum and starts decreasing without bound at high values of the transmit SNR. This confirms \textbf{\textit{Lemma 3}}. We note that the maximum of these curves decreases when the delay becomes more stringent. 

\begin{figure}[t]
   \begin{center}
      \includegraphics[width=0.5\textwidth]{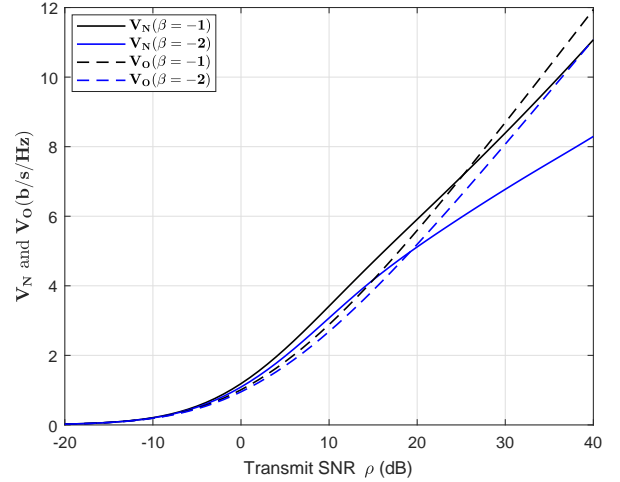}
      \caption{ $V_N$ and $V_O$ versus $\rho$, for various values of normalized delay exponent.}
      \label{fig13}
    \end{center}
  \end{figure}

 To investigate the impact of $\rho$ on the performance of the total EC for the two-user system, in Fig.\ref{fig13}, the plots for $V_N$ in NOMA and $V_O$ in OMA, versus the transmit SNR are depicted for various delay exponents. The curves demonstrate that for both NOMA and OMA, the total EC for the two users starts at the initial value of 0 and then increases with the transmit SNR, as outlined in \textbf{\textit{Lemma 4}}. When $\rho$ is very small, the total EC for the two users in NOMA, $V_N$, increases faster than $V_O$ in OMA. On the contrary, with the increase of the transmit SNR, $V_O$ becomes gradually higher than $V_N$.  At very high values of the transmit SNR, the gap between $V_N$ and $V_O$ increases further. Finally, when the delay becomes more stringent, both $V_N$ and $V_O$ decrease.
 
\begin{figure}[t]
     \begin{center}
      \includegraphics[width=0.5 \textwidth]{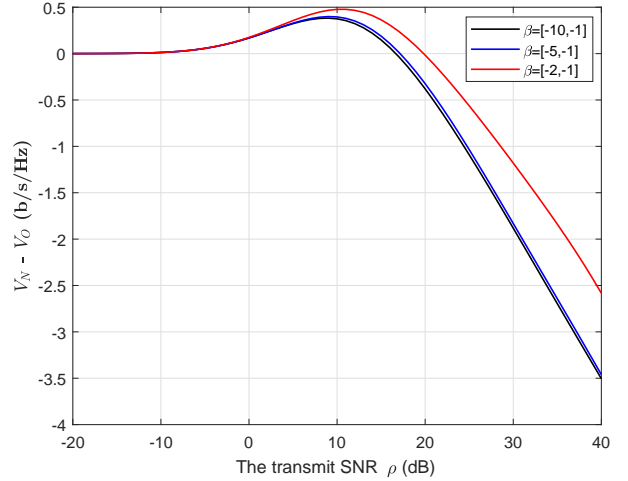}
      \caption{ $V_N$ - $V_O$ versus $\rho$ for various values of normalized delay exponent.}
      \label{fig14}
      \end{center}
  \end{figure}

\begin{figure}[t]
   \begin{center}
      \includegraphics[width=0.5\textwidth]{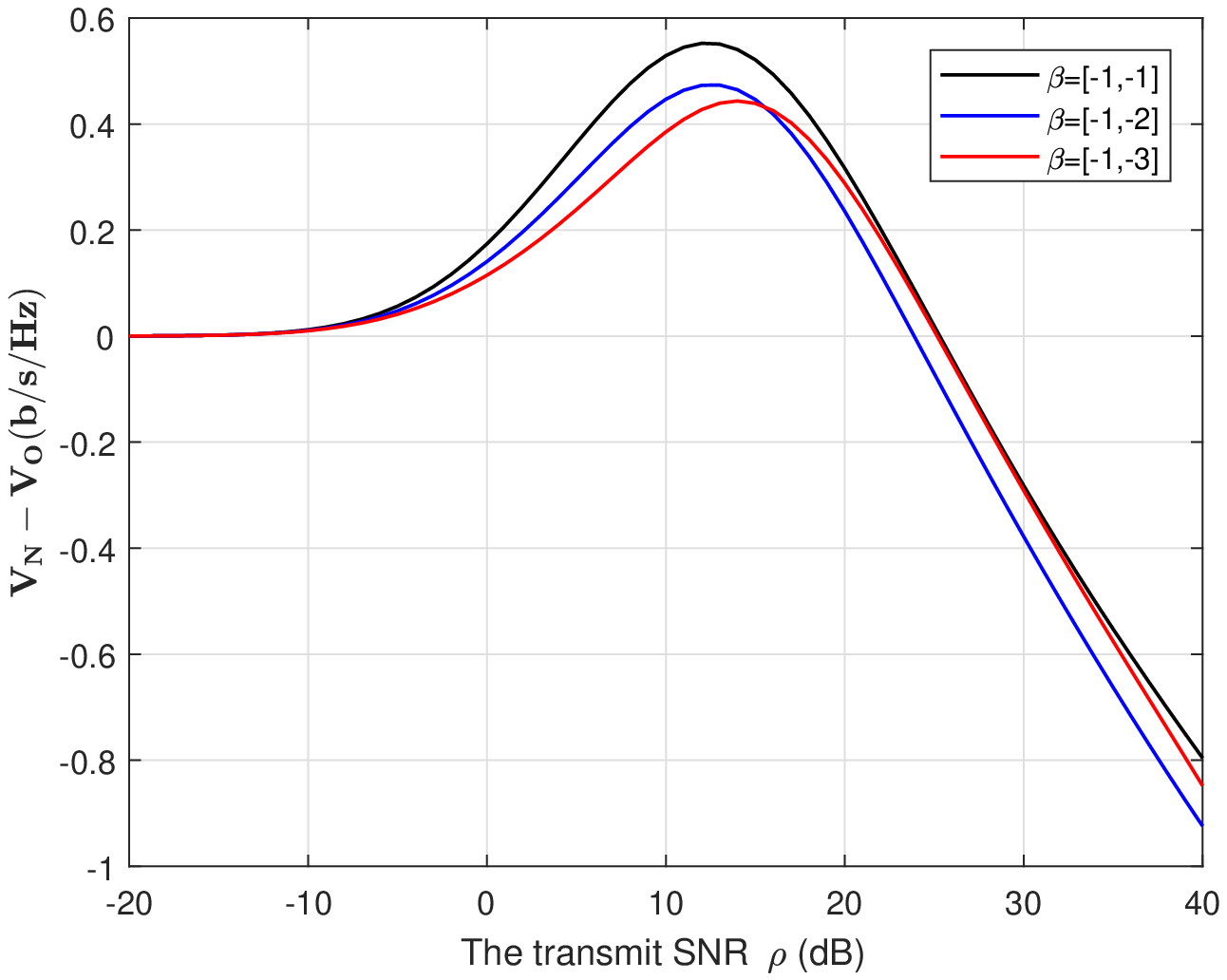}
      \caption{ $V_N$ - $V_O$ versus $\rho$ for various values of normalized delay exponent.}
      \label{fig15}
     \end{center}
  \end{figure}

 Fig.\ref{fig14} and Fig.\ref{fig15} depict $V_N-V_O$ versus $\rho$, for several values of the (negative) normalized delay exponent. In Fig.\ref{fig14}, the delay of $U_2$ is fixed, while the delay exponent of $U_1$ varies. It is shown that in that case, the smallest delay QoS (i.e., the highest negative normalized delay exponent) of $U_1$ corresponds to the highest gap in $V_N-V_O$.  On the other hand, when the delay of $U_1$ is fixed, Fig.\ref{fig15} shows that the smallest delay QoS (i.e., the highest negative normalized delay exponent) for $U_2$ corresponds to the largest gap in $V_N-V_O$. The curve of $V_N-V_O$, initially starts at zero, increases to a maximum, and returns to negative values. In the regions in which it is positive, NOMA outperforms OMA in terms of the total EC; And the opposite is true in the regions in which it is negative.

\begin{figure}[h]
    \begin{center}
     \includegraphics[width=0.5\textwidth]{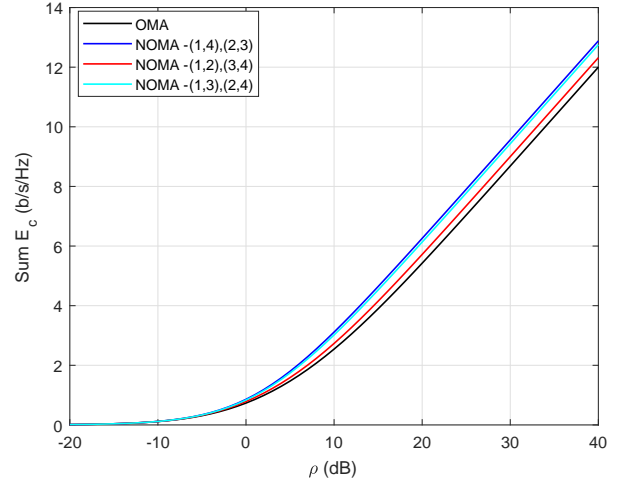}
     \includegraphics[width=0.5\textwidth]{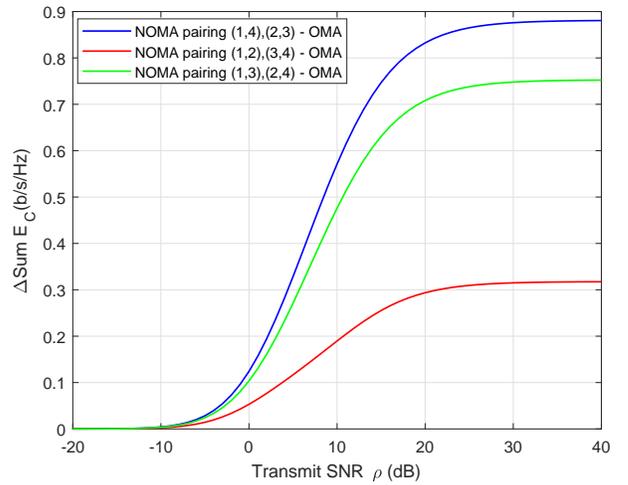}
      \caption{ (a): $E_c^{tot}$ and $\tilde{E_c}^{tot}$; (b): $E_c^{tot}$ - $\tilde{E_c}^{tot}$ versus $\rho$ for various pairing settings. $M=4$.}
     \label{fig16}
    \end{center}
\end{figure}
 \begin{figure}[h]
    \begin{center}
     \includegraphics[width=0.5\textwidth]{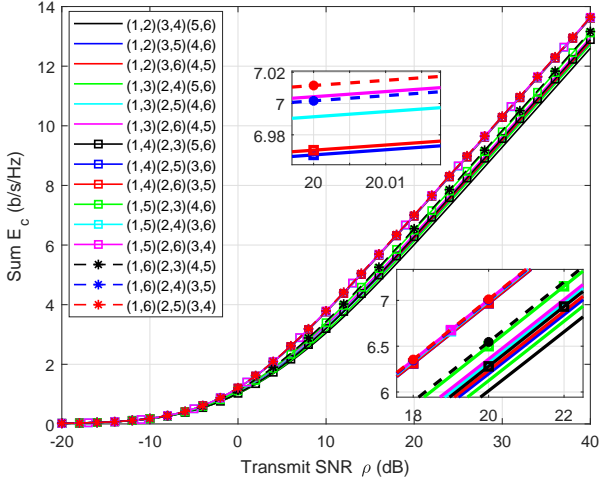}
      \caption{ $E_c^{tot}$ versus $\rho$ for various pairing settings. $M=6$.}
     \label{fig17}
    \end{center}
 \end{figure}
 \begin{figure}[h]
    \begin{center}
     \includegraphics[width=0.5\textwidth]{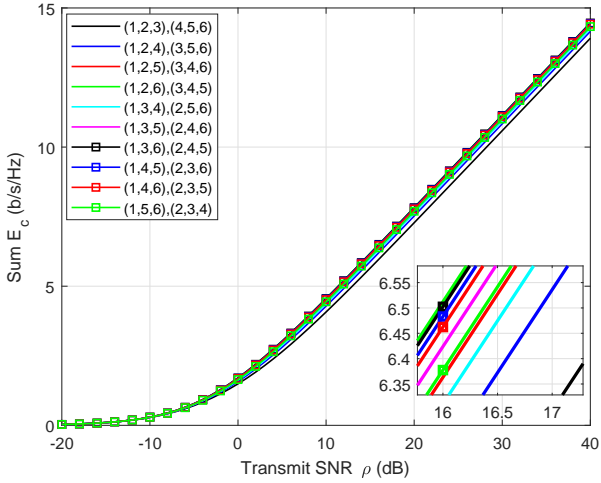}
      \caption{Sum EC versus $\rho$ for various grouping settings. $M=6$.}
     \label{fig18}
    \end{center}
 \end{figure}
\begin{figure}[h]
    \begin{center}
     \includegraphics[width=0.5\textwidth]{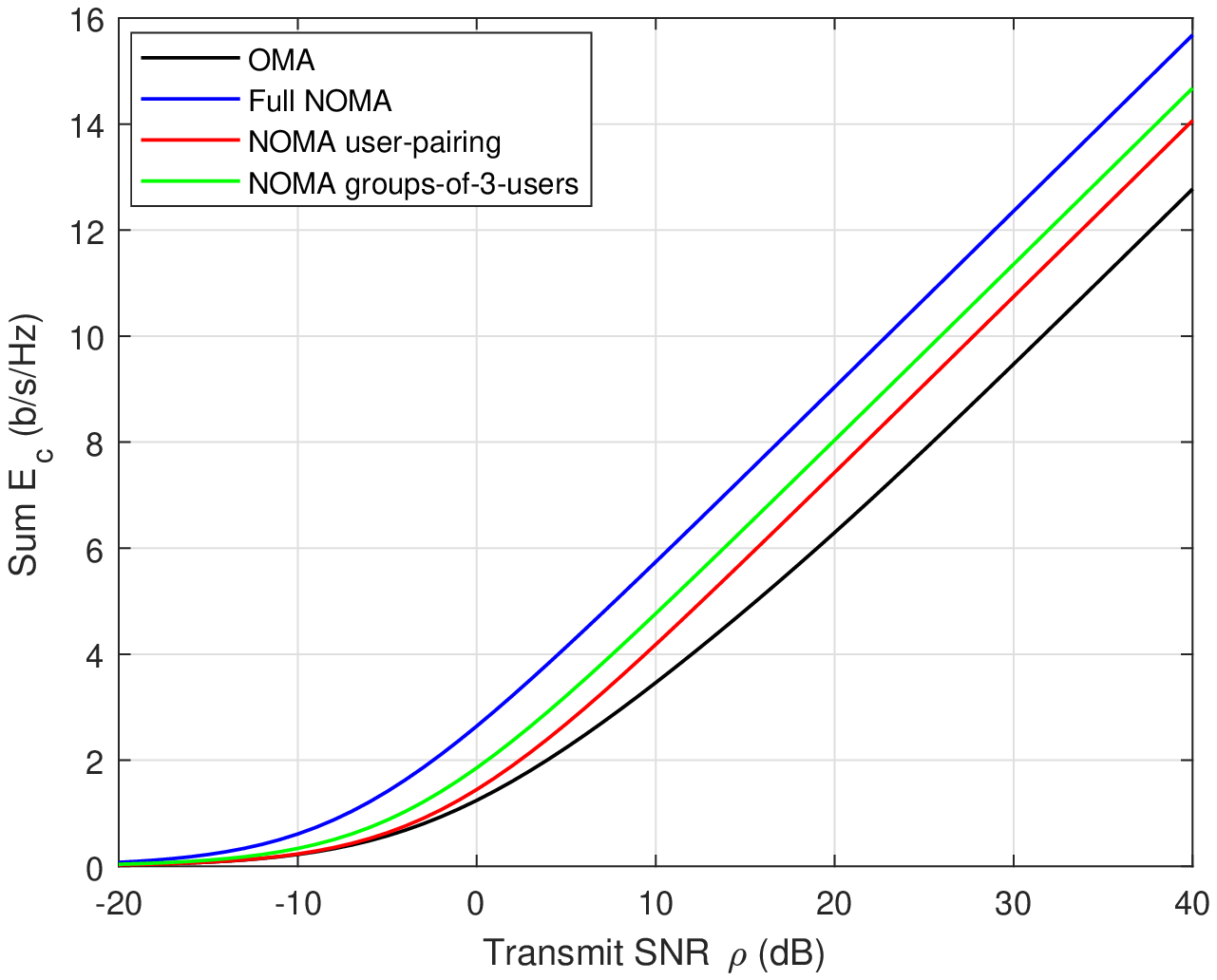}
     \includegraphics[width=0.5\textwidth]{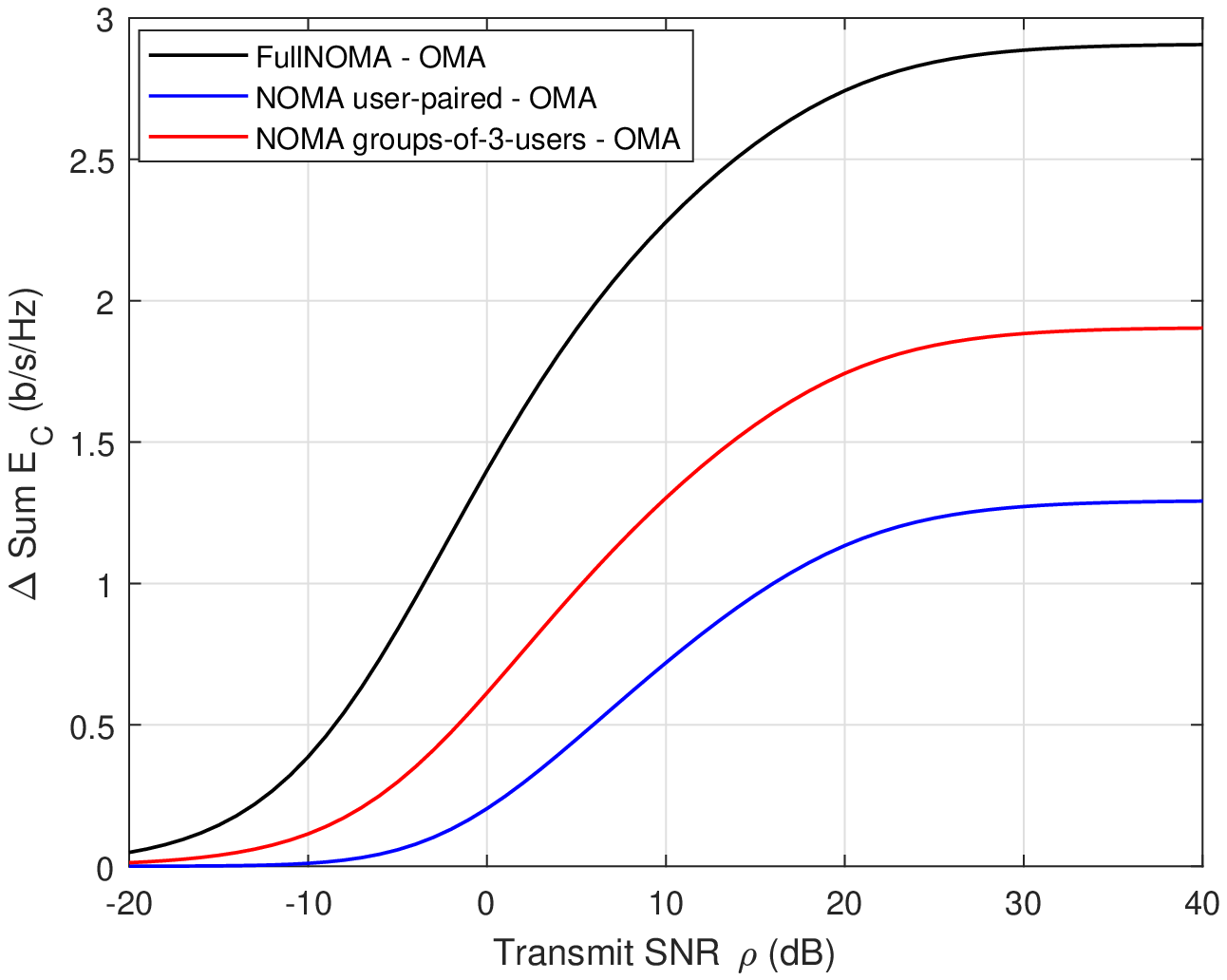}
      \caption{ Sum EC and $\Delta$Sum EC for various setting versus $\rho$. $M=6$.}
     \label{fig19}
    \end{center}
 \end{figure}
 
Next, we focus on the comparison of  multiple NOMA pairs and OMA, i.e., $E_c^{tot}$  and $\tilde{E_c}^{tot}$. Fig. \ref{fig16}-(a) depicts the curves of $\tilde{E_c}^{tot}$ and $E_c^{tot}$, versus the transmit SNR. NOMA with multiple pairs outperforms OMA. The performance gain of NOMA with multiple pairs over OMA starts at zero, increases at small values of SNR, and stabilizes at high transmit SNRs.

Fig. \ref{fig16}-(b) shows the curves of $E_c^{tot}-\tilde{E_c}^{tot}$ versus the transmit SNR, for various settings of user-pairing. Initially these start at zero at low transmit SNRs, increasing to a maximum at high values of $\rho$. This confirms \textbf{\textit{Lemma 6}}. 
Specifically, we set the total number of users $M=4$, the power coefficients allocated to both users in a NOMA pair are given as $P_{1}=0.2$ and $P_{2}=0.8$ in all the groups and the normalized delay of all users are assumed to be equal $\beta_{1,i}=\beta_{2,i}=-1$, $\left(i=1,\dots,\frac{M}{2}\right)$. The best pairing policy in the case of $M=4$ is (1,4)-(2,3). We noticed that even the worst pairing strategy outperforms OMA in terms of the total EC. 

Fig. \ref{fig17} depicts the result of the exhaustive search, done in order to find the pairing strategy which gives the highest total EC in the case of $M=6$. It appears that when these six users are divided in three groups of two users, the pairing strategy: (1,6)-(2,5)-(3,4) gives the highest total sum EC. We believe that this is due to the fact that coupling the strongest user and the weakest user produces the lowest interference when decoding. 

Fig. \ref{fig18}, on the other hand, depicts the result of the exhaustive search, of valid pairs, when all six users are divided in two groups of three users. It appears that the best pairing policy in terms of total sum EC is : (1,2,6)-(3,4,5).

Fig. \ref{fig19} depicts a comparison between full NOMA, i.e., when all users transmit in the same resource block, NOMA user-pairing, NOMA user-grouping (groups of 3 users) and OMA, for $M=6$ users. Considering the best power allocation policies in the case of user-pairing and user-grouping, it appears that full NOMA outperforms all of them in terms of the total EC, followed by NOMA with user-grouping, assuming absence of error propagation due to decoding errors. 

\section{Conclusions and Future Work}
The concept of EC enabled us to study the performance gain of NOMA over OMA in systems with statistical delay QoS constraints. First, we  investigated the EC of the uplink of a two-user NOMA network, assuming a Rayleigh block fading channel. We derived novel closed-form expressions for the ECs of the two users and provided a comparison between NOMA and OMA. The results show that, the EC of $U_1$ can surpass the EC of $U_2$, as the latter is  limited due to interference. Furthermore, we showed that the ECs of both users decrease as the delay constraints become more stringent. For both users, when the delay QoS of one of them is fixed, the smallest values of the other's delay QoS give the highest performance gap between NOMA and OMA in terms of total EC.
On the other hand, we investigated NOMA with user pairing and found the optimal pairing strategy that gave the highest EC, for $M=4$ and $M=6$. It turns out that NOMA grouping and NOMA pairing does not do better than full NOMA, but one can get close to it when users transmit with optimal power. NOMA with user pairing is interesting as it can be an alternative to mitigate interference on stronger users and reduce the impact of error propagation. These results raise questions on the possibility of switching between NOMA and OMA according to the individual users' delay constraints and transmit power. 

\section*{Appendix I}
\begin{align}
&E_c^1 = \frac{1}{\beta_1}\log_2\left(2 \int_{0}^{\infty}(1+ \rho P_1 x_1)^{\beta_1} e^{-2x_1} dx_1\right).
\end{align}
Set \(t=\rho P_1 x_1\) i.e., \(x_1=\frac{t}{\rho P_1}\) and since \(x_1:0 \rightarrow\infty \implies t:0\rightarrow\infty\), \( dx_1=\frac{1}{\rho P_1}dt\), we can get that:
\begin{align}
E_c^1 = \frac{1}{\beta_1}\log_2\left(\frac{2}{P_1 \rho} \int_{0}^{\infty}(1+ t)^{\beta_1} e^{-\frac{2t}{P_1\rho}}) dt\right).
\end{align}
Also, by setting $a=1$, $(b-a-1)=\beta_1$, $\implies$\ $b=\beta_2+2$, $z=\frac{2}{P_1 \rho}$ and denoting by \(U\left(.,.,.\right)\) the confluent hypergeometric function:
\(U(a,b,z)=\frac{1}{\Gamma(a)}\int_0^{\infty}e^{-z t} t^{a-1}(1+t)^{b-a-1}dt\), we have that:
\( \int_{0}^{\infty}(1+ t)^{\beta_1} e^{-\frac{2t}{P_1\rho}} dt=U\left(1,2+\beta_1,\frac{2}{\rho P_1}\right)\), which means that:
\begin{align}
  E_c^1 = \frac{1}{\beta_1}\log_2\left(\frac{2}{P_1 \rho}\times U\left(1,2+\beta_1,\frac{2}{\rho P_1}\right)\right).
\end{align}
For the $U_2$, we have that:
\begin{align}
  &E_c^2=\frac{1}{\beta_2}\log_2\left(\mathbb{E}\left[\left(1+ \frac{\rho P_2 x_{2}}{1+ \rho P_1 x_{1}}\right)^{\beta_2}\right]\right)\nonumber\\
  &=\!\!\frac{1}{\beta_2}\log_2\bigg(2\int_{0}^{\infty}\!\!\left(\frac{\rho P_2}{1+\rho P_1 x_1}\right)^{\beta_2}e^{-x_1}\!\!\int_{x_{1}}^{\infty}\!\!\left(\frac{1+\rho P_1 x_1}{\rho P_2}+ x_2\right)^{\beta_2}\nonumber\\
  &e^{-x_2}dx_2 dx_1 \bigg).
\end{align}
We set \(z=\frac{1+\rho P_1 x_1}{\rho P_2}+ x_2\), i.e., we have that: \(x_2=z-\frac{1+\rho P_1 x_1}{\rho P_2}\) and \(dx_2=dz\), so that \(x_2\rightarrow x_1, \implies z\rightarrow \frac{1+\rho P_1 x_1}{\rho P_2}+ x_1=\frac{1+\rho x_1}{\rho P_2}\) and \( x_2\rightarrow \infty \implies z\rightarrow \infty\).
\begin{align}
E_c^2 &=\frac{1}{\beta_2}\log_2\Bigg(2\int_{0}^{\infty}\left(\frac{\rho P_2}{1+\rho P_1 x_1}\right)^{\beta_2}e^{-x_1}\int_{\frac{1+\rho x_1}{\rho P_2}}^{\infty}z^{\beta_2}\nonumber\\
&e^{-\left(z-\frac{1+\rho P_1 x_1}{\rho P_2}\right)}dz dx_1 \Bigg)\nonumber\\
&=\frac{1}{\beta_2}\log_2\Bigg(2e^{\frac{1}{\rho P_2}}\int_{0}^{\infty}\left(\frac{\rho P_2}{1+\rho P_1 x_1}\right)^{\beta_2}e^{-x_1}e^{\frac{P_1 x_1}{ P_2}}\nonumber\\&\int_{\frac{1+\rho x_1}{\rho P_2}}^{\infty}z^{\beta_2}e^{-z}dz dx_1 \Bigg).
\end{align}
We note that:\(\int_{a}^{\infty}\frac{e^{-x}}{x^b}dx=a^{-\frac{b}{2}} e^{-\frac{a}{2}}\mathbb{W}_{-\frac{b}{2},\frac{1-b}{2}}(a)\) where $\mathbb{W}$ is the Whittaker W function. Hence, we get that:
\begin{align}
&E_c^2 =\frac{1}{\beta_2}\log_2\Bigg(2e^{\frac{1}{\rho P_2}}\int_{0}^{\infty}\left(\frac{\rho P_2}{1+\rho P_1 x_1}\right)^{\beta_2}e^{-x_1}e^{\frac{P_1 x_1}{ P_2}}\nonumber \\
&\Big[\left(\frac{1+\rho x_1}{\rho P_2}\right)^{\frac{\beta_2}{2}}e^{-\frac{1+\rho x_1}{2 \rho P_2}}\mathbb{W}_{\frac{\beta_2}{2},\frac{1+\beta_2}{2}}\left(\frac{1+\rho x_1}{\rho P_2}\right) \Big] dx_1 \Bigg)\nonumber\\
&=\frac{1}{\beta_2}\log_2\Bigg(2\left(\rho P_2\right)^{\frac{\beta_2}{2}} e^{\frac{1}{2 \rho P_2}}\int_{0}^{\infty}\left(1+\rho P_1 x_1\right)^{-\beta_2} \nonumber\\
&\left(1+\rho x_1\right)^{\frac{\beta_2}{2}}e^{\frac{(2P_1-2 P_2-1) x_1}{2 P_2}}\Big[\mathbb{W}_{\frac{\beta_2}{2},\frac{1+\beta_2}{2}}\left(\frac{1+\rho x_1}{\rho P_2}\right) \Big] dx_1 \Bigg).
\end{align}
Note that \(\mathbb{W}_{u-\frac{1}{2},u}(z)=e^{\frac{1}{2}z} z^{\frac{1}{2}-u}\Gamma(2 u,z)\), so that we have
\(\mathbb{W}_{\frac{\beta_2}{2},\frac{1+\beta_2}{2}}\left(\frac{1+\rho x_1}{\rho P_2}\right)=e^{\frac{1+\rho x_1}{2 \rho P_2}}\left(\frac{1+\rho x_1}{\rho P_2}\right)^{-\frac{\beta_2}{2}}\Gamma\left(1+\beta_2,\frac{1+\rho x_1}{\rho P_2}\right)\).

By substituting it in $E_c^2$, we have that: 
\begin{align}
&E_c^2 
=\frac{1}{\beta_2}\log_2\Bigg(2(\rho P_2)^{\beta_2} e^{\frac{1}{\rho P_2}}\int_{0}^{\infty}(1+\rho P_1 x_1)^{-\beta_2} e^{\frac{(P_1-P_2) x_1}{P_2}}\nonumber\\ &\times\Gamma\left(1+\beta_2,\frac{1+\rho x_1}{\rho P_2}\right) dx_1 \Bigg).
\end{align}
To continue we set $\frac{1+\rho x_1}{\rho P_2}=y$, i.e., $x_1=P_2 y-\frac{1}{\rho}$, and $dx_1=P_2dy$.
$x_1\rightarrow0$ $\implies$ $y\rightarrow\frac{1}{\rho P_2}$ and
$x_1\rightarrow\infty$ $\implies$ $y\rightarrow\infty$. 
Recall that without loss of generality we have set $P_1+P_2=1$. Then we get that
\begin{align}
&E_c^2=\frac{1}{\beta_2}\log_2\Bigg(2(\rho P_2)^{\beta_2} e^{\frac{1}{\rho P_2}}\int_{0}^{\infty}(1+\rho P_1 x_1)^{-\beta_2}\nonumber\\
& \times e^{\frac{(P_1-P_2) x_1}{P_2}}\Big[\Gamma\left(1+\beta_2,\frac{1+\rho x_1}{\rho P_2}\right) \Big] dx_1 \Bigg)\nonumber\\
&=\frac{1}{\beta_2}\log_2\bigg(2 P_2 (\rho P_2)^{\beta_2} e^{\frac{1}{\rho P_2}}e^{-\frac{(P_1-P_2)}{\rho P_2}}\nonumber\\
&\times\int_{\frac{1}{\rho P_2}}^{\infty}P_2^{-\beta_2}(1+\rho P_1 y)^{-\beta_2} e^{(P_1-P_2)y}\Gamma(1+\beta_2,y) dy \bigg).
\end{align}
Using binomial expansion we have $(1+\rho P_1 y)^{-\beta_2}=\sum_{j=0}^{-\beta_2}\binom{-\beta_2}{j}(\rho P_1 y)^{j}$ when $\beta_2$ is integer, otherwise we use $\lfloor\beta_2\rfloor$. And, using Taylor series expansion we have that $ e^{(P_1-P_2)y}= e^{-(P_2-P_1)y}=\sum_{k=0}^{\infty}\frac{(-1)^k (P_2-P_1)^k}{k!}y^k$, which converges.
\begin{align}
E_c^2&=\frac{1}{\beta_2}\log_2\bigg(2 P_2^{1-\beta_2}(\rho P_2)^{\beta_2} e^{\frac{1}{\rho P_2}}e^{-\frac{(P_1-P_2)}{\rho P_2}}\nonumber\\&\times\int_{\frac{1}{\rho P_2}}^{\infty}(1+\rho P_1 y)^{-\beta_2} e^{(P_1-P_2)y}\Gamma(1+\beta_2,y)dy \bigg)\nonumber\\
&=\frac{1}{\beta_2}\log_2\bigg(2 P_2^{1-\beta_2}(\rho P_2)^{\beta_2} e^{\frac{1}{\rho P_2}}e^{-\frac{(P_1-P_2)}{\rho P_2}}\nonumber\\&\times\sum_{j=0}^{-\beta_2}\binom{-\beta_2}{j}(\rho P_1)^{j}\times \sum_{k=0}^{\infty}\frac{(-1)^k (P_2-P_1)^k}{k!} \nonumber\\&\times\int_{\frac{1}{\rho P_2}}^{\infty}y^{j+k} \Gamma\left(1+\beta_2,y\right) dy \bigg).
\label{eqf4}
\end{align}
Note that
\begin{equation}
  \int_{c}^{\infty}\!\!y^{b} \Gamma\left(A,z\right)dz\!\!  =\frac{1}{1+b}\bigg(-c^{1+b}\Gamma\left(A,c\right)+\Gamma\left(1+A+b,c\right)\bigg)\nonumber
\end{equation}
i.e.,
\begin{align}
&\int_{\frac{1}{\rho P_2}}^{\infty}y^{j+k} \Gamma\left(1+\beta_2,y\right) dy=\frac{1}{1+j+k}\nonumber\\
\!&\!\!\times
\!\!\bigg(\!\!-(\rho P_2)^{-1-j-k}\Gamma(1+\beta_2,\frac{1}{\rho P_2})\!+\!\Gamma(2+\beta_2+j+k,\frac{1}{\rho P_2})\!\!\bigg).
\label{lastt}
\end{align}
Finally, by inserting \eqref{lastt} in \eqref{eqf4} we obtain \eqref{eq:EC2}.

\section*{Appendix II}
By inserting $\rho\rightarrow0$ into \eqref{eq:EC1} and \eqref{eq:EC2}, we get 1) of \textbf{\textit{Lemma 1}}, i.e.,
\begin{equation}
   \lim\limits_{\rho\rightarrow0}(E_c^1-\widetilde{E}_c^{1})=\frac{1}{\beta_1}\log_2\left(\frac{\mathbb{E}\left[\left(1+\rho P_1 |h_1|^2\right)^{\beta_2}\right]}{\mathbb{E}\left[\left(1+2 \rho P_1 |h_1|^2\right)^{\frac{\beta_2}{2}}\right]}\right)=0 \nonumber,
\end{equation}
\begin{equation}
   \lim\limits_{\rho\rightarrow0}(E_c^2-\widetilde{E}_c^{2})=\frac{1}{\beta_2}\log_2\left(\frac{\mathbb{E}\Big[\left(1+\frac{\rho P_2 |h_2|^2}{1+\rho P_1 |h_1|^2}\right)^{\beta_2}\Big]}{\mathbb{E}\left[\left(1+2\rho P_2 |h_1|^2\right)^{\frac{\beta_2}{2}}\right]}\right)=0 \nonumber.
\end{equation}

In the same way, by inserting $\rho\rightarrow \infty$ into \eqref{eq:EC1} and \eqref{eq:EC2}, we get 2) in \textbf{\textit{Lemma 1}}, given below.

\begin{align}
\lim\limits_{\rho\rightarrow \infty}E_c^2&= \frac{1}{\beta_2}\log_2\left(\mathbb{E}\left[\left(1+\frac{P_2 |h_2|^2}{P_1 |h_1|^2}\right)^{\beta_2}\right]\right) \nonumber,\\
 \lim\limits_{\rho\rightarrow \infty}(E_c^1-\widetilde{E}_c^{1})&=\frac{1}{\beta_1}\log_2\left((\rho P_1)^{\frac{\beta_1}{2}} \frac{\mathbb{E}\left[(\frac{1}{\rho P_1}+ |h_1|^2)^{\beta_2}\right]}{\mathbb{E}\left[(\frac{1}{\rho P_1}+ 2 |h_1|^2)^{\frac{\beta_2}{2}}\right]}\right)\nonumber\\&=\infty \nonumber,\\
 \lim\limits_{\rho\rightarrow \infty}(E_c^2-\widetilde{E}_c^{2})&=\frac{1}{\beta_2}\log_2\left(\frac{\mathbb{E}\left[\left(\frac{\frac{1}{\rho}+ P_1 |h_1|^2+ P_2 |h_2|^2}{\frac{1}{\rho}+ P_1 |h_1|^2}\right)^{\beta_2}\right]}{\rho^{\frac{\beta_2}{2}}\mathbb{E}\left[\left(\frac{1}{\rho}+ 2 P_2 |h_2|^2\right)^{\frac{\beta_2}{2}}\right]}\right)\nonumber\\&= -\infty \nonumber.
\end{align}

\section*{Appendix III}
To analyze the trends of $E_c^1$ and $\widetilde{E}_c^{1}$ with respect to $\rho$, we start with

\begin{align}
    \frac{\partial E_c^1}{\partial\rho}&=\frac{1}{\beta_1 \ln2}\frac{\Big(\mathbb{E}[(1+\rho P_1 |h_1|^2)^{\beta_1}] \Big)'}{\mathbb{E}[(1+\rho P_1 |h_1|^2)^{\beta_1}]}\nonumber\\
    &=\frac{P_1}{\ln2} \frac{\mathbb{E}[|h_1|^2 (1+\rho P_1 |h_1|^2)^{\beta_1-1}]}{\mathbb{E}[(1+\rho P_1 |h_1|^2)^{\beta_1}]}\ge 0.
\end{align}
Similarly, for $U_1$ in OMA we have that
\begin{align}
    \frac{\partial \widetilde{E}_c^{1}}{\partial\rho}&=\frac{1}{\beta_1\ln2}\frac{\Big(\mathbb{E}[(1+2 \rho P_1 |h_1|^2)^{\frac{\beta_1}{2}}] \Big)'}{\mathbb{E}[(1+2 \rho P_1 |h_1|^2)^{\frac{\beta_1}{2}}]}\nonumber\\
    &=\frac{P_1}{\ln2} \frac{\mathbb{E}[|h_1|^2 (1+2 \rho P_1 |h_1|^2)^{\frac{\beta_1}{2}-1}]}{\mathbb{E}[(1+2 \rho P_1 |h_1|^2)^{\frac{\beta_1}{2}}]} \ge 0.
\end{align}
Then, we get that
\begin{align}
    \frac{\partial (E_c^1-\widetilde{E}_c^{1})}{\partial\rho}&=\frac{P_1}{\ln2} \frac{\mathbb{E}[|h_1|^2 (1+\rho P_1 |h_1|^2)^{\beta_1-1}]}{\mathbb{E}[(1+\rho P_1 |h_1|^2)^{\beta_1}]}\nonumber\\
    &-\frac{P_1}{\ln2} \frac{\mathbb{E}[|h_1|^2 (1+2\rho P_1|h_1|^2)^{\frac{\beta_1}{2}-1}]}{\mathbb{E}[(1+2\rho P_1|h_1|^2)^{\frac{\beta_1}{2}}]}.
\end{align}
and
$\lim\limits_{\rho\rightarrow0}(\frac{\partial (E_c^1-\widetilde{E}_c^{1})}{\partial\rho})=\frac{(P_1-P_1)}{\ln 2}\mathbb{E}[|h_1|^2] =0$. When $\rho>>1$, we have that
\begin{align}
\frac{\partial (E_c^1-\widetilde{E}_c^{1})}{\partial\rho})&=\frac{P_1}{\rho \ln2} \frac{\mathbb{E}[|h_1|^2 (P_1 |h_1|^2)^{\beta_1-1}]}{\mathbb{E}[(P_1 |h_1|^2)^{\beta_1}]}\nonumber\\
&-\frac{P_1}{\rho \ln2} \frac{\mathbb{E}[|h_1|^2 (2 P_1 |h_1|^2)^{\frac{\beta_1}{2}-1}]}{\mathbb{E}[(2 P_1 |h_1|^2)^{\frac{\beta_1}{2}}]}\nonumber\\
&=\frac{1}{2\rho \ln2}\ge0.
\end{align}
When $\rho\rightarrow\infty$, this term approaches 0.

\section*{Appendix IV}
\begin{align}
E_c^2=\frac{1}{\beta_2}\log_2\left(\mathbb{E}\Big[\left(1+ \frac{\rho P_2 |h_2|^2}{1+ \rho P_1 |h_1|^2}\right)^{\beta_2}\Big]\right).
\end{align}
And
\begin{align}
    \frac{\partial E_c^2}{\partial\rho}&=\frac{1}{\beta_2 \ln2}\frac{\left(\mathbb{E}\Big[\left(1+ \frac{\rho P_2 |h_2|^2}{1+  \rho P_1 |h_1|^2}\right)^{\beta_2}\Big]\right)'}{\mathbb{E}\Big[\left(1+ \frac{\rho P_2 |h_2|^2}{1+  \rho P_1 |h_1|^2}\right)^{\beta_2}\Big]}\nonumber\\
     \!\!\!\!  \!\!\!\! &= \!\! \frac{1}{ \ln2}\frac{\mathbb{E}\Big[\frac{P_2 |h_2|^2}{(1+\rho P_1 |h_1|^2)^2}\left(1+ \frac{\rho P_2 |h_2|^2}{1+  \rho P_1 |h_1|^2}\right)^{\beta_2-1}\Big]}{\mathbb{E}\Big[\left(1+ \frac{\rho P_2 |h_2|^2}{1+  \rho P_1 |h_1|^2}\right)^{\beta_2}\Big]}\ge0.
\end{align} 
In the same way, for the $U_2$ in OMA, we have that:
\begin{align}
    \frac{\partial \widetilde{E}_c^{2}}{\partial\rho}&=\frac{1}{\beta_2\ln2}\frac{\Big(\mathbb{E}[(1+2\rho P_2 |h_2|^2)^{\frac{\beta_2}{2}}] \Big)'}{\mathbb{E}[(1+2 \rho P_2 |h_2|^2)^{\frac{\beta_2}{2}}]}\nonumber\\
    &=\frac{P_2}{\ln2} \frac{\mathbb{E}[|h_2|^2 (1+2\rho P_2 |h_2|^2)^{\frac{\beta_2}{2}-1}]}{\mathbb{E}[(1+2 \rho P_2 |h_2|^2)^{\frac{\beta_2}{2}}]} \ge 0,
\end{align}
and
\begin{align}
\frac{\partial (E_c^2-\widetilde{E}_c^{2})}{\partial\rho}&=\frac{1}{ \ln2}\frac{\mathbb{E}\Big[\frac{P_2 |h_2|^2}{(1+\rho P_1 |h_1|^2)^2}\left(1+ \frac{\rho P_2 |h_2|^2}{1+  \rho P_1 |h_1|^2}\right)^{\beta_2-1}\Big]}{\mathbb{E}\Big[\left(1+ \frac{\rho P_2 |h_2|^2}{1+  \rho P_1 |h_1|^2}\right)^{\beta_2}\Big]}\nonumber\\
&-\frac{P_2}{\ln2} \frac{\mathbb{E}[|h_2|^2 (1+2\rho P_2|h_2|^2)^{\frac{\beta_2}{2}-1}]}{\mathbb{E}[(1+2 \rho P_2 |h_2|^2)^{\frac{\beta_2}{2}}]}.
\end{align}
When $\rho\rightarrow0$, we have that
$\lim\limits_{\rho\rightarrow0}(\frac{\partial (E_c^2-\widetilde{E}_c^{2})}{\partial\rho})=0$.
When $\rho$ is very large,
\begin{align}
&\frac{\partial (E_c^2-\widetilde{E}_c^{2})}{\partial\rho}=\frac{\mathbb{E}\Big[\frac{P_2 |h_2|^2}{\rho^2 (\frac{1}{\rho}+ P_1 |h_1|^2)^2}(1+\frac{\rho}{\rho} \frac{(P_2 |h_2|^2)}{(\frac{1}{\rho}+ P_1 |h_1|^2)})^{\beta_2-1}\Big]}{\ln2 \mathbb{E}\Big[(1+ \frac{\rho}{\rho}\frac{P_2 |h_2|^2}{(\frac{1}{\rho}+ P_1 |h_1|^2)})^{\beta_2}\Big]}\nonumber\\ 
&-\frac{P_2}{\ln2}\frac{1}{\rho} \frac{\mathbb{E}[|h_2|^2 (\frac{1}{\rho}+2 P_2 |h_2|^2)^{\frac{\beta_2}{2}-1}]}{\mathbb{E}[(\frac{1}{\rho}+ 2 P_2|h_2|^2)^{\frac{\beta_2}{2}}]}\nonumber\\
&=\frac{P_2}{\rho^2 P_{1}^2 \ln2}\frac{\mathbb{E}\Big[\frac{|h_2|^2}{(|h_1|^2)^2}\left(1+\frac{P_2 |h_2|^2}{P_1 |h_1|^2}\right)^{\beta_2-1}\Big]}{ \mathbb{E}\Big[\left(1+\frac{P_2 |h_2|^2}{P_1 |h_1|^2}\right)^{\beta_2}\Big]} -\frac{1}{2 \ln2}\frac{1}{\rho} \nonumber\\
&=\frac{\frac{P_2}{P_{1}^2\ln2} A-\frac{1}{2 \ln2}\rho}{\rho^2},
\end{align}
where $A=\frac{\mathbb{E}\Big[\frac{|h_2|^2}{(|h_1|^2)^2}\left(1+\frac{P_2 |h_2|^2}{P_1 |h_1|^2}\right)^{\beta_2-1}\Big]}{\mathbb{E}\Big[\left(1+\frac{P_2 |h_2|^2}{P_1 |h_1|^2}\right)^{\beta_2}\Big]}$, unrelated to $\rho$.
And it gradually approaches 0 when $\rho\rightarrow\infty$.

\section*{Appendix V}
Note that $V_N=E_c^1+E_c^{2}$. By using \textbf{\textit{Lemma 1}}, we have that
$ \lim\limits_{\rho\rightarrow0}(V_N)=0$ and $ \lim\limits_{\rho\rightarrow\infty}(V_N)=\infty$. Then, we get that,
\begin{align}
\frac{\partial V_N}{\partial\rho}&=\frac{\partial (E_c^1+E_c^{2})}{\partial\rho}=\frac{P_1}{\ln2} \frac{\mathbb{E}[|h_1|^2 (1+\rho P_1 |h_1|^2)^{\beta_1-1}]}{\mathbb{E}[(1+\rho P_1 |h_1|^2)^{\beta_1}]}\nonumber\\
&+\frac{1}{ \ln2}\frac{\mathbb{E}\Big[\frac{P_2 |h_2|^2}{(1+\rho P_1 |h_1|^2)^2}\left(1+ \frac{\rho P_2 |h_2|^2}{1+  \rho P_1 |h_1|^2}\right)^{\beta_2-1}\Big]}{\mathbb{E}\Big[\left(1+ \frac{\rho P_2 |h_2|^2}{1+  \rho P_1 |h_1|^2}\right)^{\beta_2}\Big]}\geq 0.
\end{align}

When $\rho\rightarrow 0$, we have that
$ \lim\limits_{\rho\rightarrow 0}(\frac{\partial V_N}{\partial\rho})=\frac{P_1}{\ln2}\mathbb{E}[|h_1|^2]+\frac{P_2}{ \ln2}\mathbb{E}[|h_2|^2]$. 
When $\rho\rightarrow\infty$, we get that
\begin{equation}
\lim\limits_{\rho\rightarrow\infty}\frac{\partial V_N}{\partial\rho}=\frac{1}{\rho \ln2}+\frac{\mathbb{E}\Big[\frac{P_2 |h_2|^2}{(P_1 |h_1|^2)^2}\left(1+ \frac{P_2 |h_2|^2}{P_1 |h_1|^2}\right)^{\beta_2-1}\Big]}{\rho^2 \ln2\mathbb{E}\Big[\left(1+ \frac{P_2 |h_2|^2}{ P_1 |h_1|^2}\right)^{\beta_2}\Big]} \nonumber  \!\!=\!\!0.
\end{equation} 

For $V_O$ in the case of OMA, we note that
$V_O=\widetilde{E}_c^{1}+\widetilde{E}_c^{2}$. By using \textbf{\textit{Lemma 1}}, we have 
$ \lim\limits_{\rho\rightarrow0}(V_0)=0$ and $ \lim\limits_{\rho\rightarrow\infty}(V_0)=\infty$.
Then,
\begin{align}
\frac{\partial V_0}{\partial\rho}&=\frac{\partial (\widetilde{E}_c^{1}+\widetilde{E}_c^{2})}{\partial\rho}=\frac{P_1}{\ln2} \frac{\mathbb{E}[|h_1|^2 (1+2\rho P_1 |h_1|^2)^{\frac{\beta_1}{2}-1}]}{\mathbb{E}[(1+2\rho P_1|h_1|^2)^{\frac{\beta_1}{2}}]}\nonumber \\
&+\frac{P_2}{\ln2} \frac{\mathbb{E}[|h_2|^2 (1+2\rho P_2|h_2|^2)^{\frac{\beta_2}{2}-1}]}{\mathbb{E}[(1+2\rho P_2 |h_2|^2)^{\frac{\beta_2}{2}}]} \geq 0.
\end{align}
When $\rho\rightarrow0$, we have that $ \lim\limits_{\rho\rightarrow 0}(\frac{\partial V_O}{\partial\rho})=\frac{P_1}{\ln2}\mathbb{E}[|h_1|^2]+\frac{P_2}{\ln2}\mathbb{E}[|h_2|^2]$. 
When $\rho\rightarrow\infty$, we have that $\lim\limits_{\rho\rightarrow\infty}(\frac{\partial V_O}{\partial\rho})= \lim\limits_{\rho\rightarrow\infty}(\frac{1}{2 \rho \ln2}+\frac{1}{2 \rho \ln2})=\lim\limits_{\rho\rightarrow\infty}(\frac{1}{ \rho \ln2})$,
which equals to $0$. 

\section*{Appendix VI}
We have
\begin{align}
   E_c^2 &= \frac{1}{\beta_2}\log_2\left(\mathbb{E}\Big[\left(1+ \frac{\rho P_2 |h_2|^2}{1+ \rho P_1 |h_1|^2}\right)^{\beta_2}\Big]\right)\nonumber\\
   &=-\frac{1}{\theta_2 T_f B}\left(\mathbb{E}\Big[-\frac{\theta_2 T_f B}{\ln2} \ln\left(1+ \frac{\rho P_2 |h_2|^2}{1+ \rho P_1 |h_1|^2}\right)\Big]\right).
\end{align}
When $\theta_2\rightarrow0$, we get an indeterminate form. By applying the L'Hopital's rule one can get 
\begin{align}
   E_c^2 &=-\frac{1}{T_f B}\left(\mathbb{E}\left[-\frac{ T_f B}{\ln2} \ln\Big(1+ \frac{\rho P_2 |h_2|^2}{1+ \rho P_1 |h_1|^2}\Big)\right]\right)\nonumber\\
   &=\mathbb{E}\left[\log_2\left(1+ \frac{\rho P_2 |h_2|^2}{1+ \rho P_1 |h_1|^2}\right)\right].
\end{align} 
Hence, we get that
\begin{align}
   \lim\limits_{\theta_2\rightarrow0}E_c^2 =\mathbb{E}\left[\log_2\left(1+ \frac{\rho P_2 |h_2|^2}{1+ \rho P_1 |h_1|^2}\right)\right],\nonumber
\end{align} 
which equals to $\mathbb{E}[R_2]$, the ergodic capacity.
 
Proceeding in the same way, one can find
\begin{align*}
&\lim\limits_{\theta_1\rightarrow0}E_c^1=\mathbb{E}\Big[\log_2\Big(1+\rho P_1 |h_1|^2\Big)\Big]=\mathbb{E}[R_1],\nonumber\\
&\lim\limits_{\theta_1\rightarrow0}\widetilde{E}_c^{1}=\mathbb{E}\Big[\frac{1}{2}\log_2\Big(1+2\rho P_1 |h_1|^2\Big)\Big]=\mathbb{E}[\widetilde{R}_{1}],\nonumber\\
&\lim\limits_{\theta_2\rightarrow0}\widetilde{E}_c^{2}=\mathbb{E}\Big[\frac{1}{2}\log_2\Big(1+2\rho P_2 |h_2|^2\Big)\Big]=\mathbb{E}[\widetilde{R}_{2}],\nonumber\\
\\
&\lim\limits_{\theta_1\rightarrow0}(E_c^1-\widetilde{E}_c^{1})=\mathbb{E}[R_1]-\mathbb{E}[\widetilde{R}_{1}],\nonumber\\
&\lim\limits_{\theta_2\rightarrow0}(E_c^2-\widetilde{E}_c^{2})=\mathbb{E}[R_2]-\mathbb{E}[\widetilde{R}_{2}].\nonumber
\end{align*}
To look further the impact of the transmit SNR $\rho$ on the $EC$ considering delay-unconstrained user:
\begin{align}
   \lim\limits_{\substack{\theta_1\rightarrow0\\\rho\rightarrow\infty}}E_c^1 =\lim\limits_{\substack{\rho\rightarrow\infty}}\mathbb{E}\Big[\log_2\Big(1+\rho P_1 |h_1|^2\Big)\Big]=\infty,\nonumber
\end{align} 
We also have that
\begin{align}
   \lim\limits_{\substack{\theta_2\rightarrow0\\\rho\rightarrow\infty}}E_c^2 &=\lim\limits_{\substack{\rho\rightarrow\infty}}\mathbb{E}\left[\log_2\left(1+ \frac{\rho P_2 |h_2|^2}{1+ \rho P_1 |h_1|^2}\right)\right]\nonumber\\
   &=\mathbb{E}\left[\log_2\left(1+ \frac{P_2 |h_2|^2}{P_1 |h_1|^2}\right)\right].\nonumber
\end{align}
Similarly, we have for OMA
\begin{align*}
&\lim\limits_{\substack{\theta_1\rightarrow0\\\rho\rightarrow\infty}}\widetilde{E}_c^{1}=\lim\limits_{\substack{\rho\rightarrow\infty}}\mathbb{E}\Big[\frac{1}{2}\log_2\left(1+2\rho P_1|h_1|^2\right)\Big]=\infty,\\
&\lim\limits_{\substack{\theta_2\rightarrow0\\\rho\rightarrow\infty}}\widetilde{E}_c^{2}=\lim\limits_{\substack{\rho\rightarrow\infty}}\mathbb{E}\Big[\frac{1}{2}\log_2\Big(1+2 \rho P_2|h_2|^2\Big)\Big]=\infty.
\end{align*}
Therefore, we have that
\begin{align*}
   \lim\limits_{\substack{\theta_1\rightarrow0\\
   \rho\rightarrow\infty}}\left(E_c^1-\widetilde{E}_c^{1}\right) &=\lim\limits_{\substack{\rho\rightarrow\infty}}\left(\mathbb{E}\Big[\log2\left(\frac{1+\rho P_1 |h_1|^2}{(1+2\rho P_1 |h_1|^2)^{\frac{1}{2}}}\right)\Big]\right)\\
   &= \lim\limits_{\substack{\rho\rightarrow\infty}}\left(\mathbb{E}\Big[\log2\left(\sqrt{\frac{\rho P_1|h_1|^2}{2}}\right)\Big]\right)=\infty.\\
   \lim\limits_{\substack{\theta_2\rightarrow0\\\rho\rightarrow\infty}}\left(E_c^2-\widetilde{E}_c^{2}\right) &=-\infty.
\end{align*}
\section*{Appendix VII}

Using the \textbf{\textit{Lemma 1}}, when $\rho\rightarrow0$, we can show that $E_c^{1,i}-\widetilde{E}_c^{1,i}\rightarrow0$ and $E_c^{2,i}-\widetilde{E}_c^{2,i}\rightarrow0$. Then $E_c^{tot}-\tilde{E_c}^{tot}\rightarrow0$, since 
$E_c^{tot}-\tilde{E_c}^{tot}=\sum_{i=1}^{\frac{M}{2}}(E_c^{1,i}+E_c^{2,i}-\widetilde{E}_c^{1,i}-\widetilde{E}_c^{2,i})$, we get 
\begin{align*}
     \lim\limits_{\rho\rightarrow0}\left(E_c^{tot}-\tilde{E_c}^{tot}\right)=0.
\end{align*}
On the other side, when $\rho\rightarrow\infty$,
\begin{align}
    &E_c^{tot}-\tilde{E_c}^{tot}=\sum_{i=1}^{\frac{M}{2}}\Bigg(\frac{1}{\beta_{1,i}}\log_2\left(\frac{\mathbb{E}\Big[(1+ \rho P_{1,i}|h_{1,i}|^2)^{\frac{2 \beta_{1,i}}{M}}\Big]}{\mathbb{E}\Big[(1+ 2 \rho P_{1,i} |h_{1,i}|^2)^{\frac{\beta_{1,i}}{M}}\Big]}\right)\nonumber\\
    &+\frac{1}{\beta_{2,i}}\log_2\left(\frac{\mathbb{E}\Big[\left(1+ \frac{\rho P_{2,i}|h_{2,i}|^2}{1+ \rho P_{1,i}|h_{1,i}|^2}\right)^{\frac{2 \beta_{2,i}}{M}}\Big]}{\mathbb{E}\Big[(1+2 \rho P_{2,i} |h_{2,i}|^2)^{\frac{\beta_{2,i}}{M}}\Big]}\right)\Bigg)\nonumber \\
    &=\!\!\sum_{i=1}^{\frac{M}{2}}\!\!\Bigg(\frac{1}{\beta_{1,i}}\!\log_2\left(\rho^{\frac{\beta_{1,i}}{M}}\frac{\mathbb{E}\Big[(\frac{1}{\rho}+ P_{1,i}|h_{1,i}|^2)^{\frac{2 \beta_{1,i}}{M}}\Big]}{\mathbb{E}\Big[(\frac{1}{\rho}+ 2 P_{1,i} |h_{1,i}|^2)^{\frac{\beta_{1,i}}{M}}\Big]}\!\!\right)\nonumber\\
    &\!\!+\frac{1}{\beta_{2,i}}\log_2\left(\rho^{-\frac{\beta_{2,i}}{M}}\frac{\mathbb{E}\Big[\left(1+ \frac{P_{2,i}|h_{2,i}|^2}{\frac{1}{\rho}+  P_{1,i}|h_{1,i}|^2}\right)^{\frac{2 \beta_{2,i}}{M}}\Big]}{\mathbb{E}\Big[(\frac{1}{\rho}+ 2 P_{2,i}|h_{2,i}|^2)^{\frac{\beta_{2,i}}{M}}\Big]}\!\!\right)\Bigg).
\end{align}
Then,
\begin{align}
E_c^{tot}-\tilde{E_c}^{tot}&=\nonumber\\
&\sum_{i=1}^{\frac{M}{2}}\Bigg(\frac{1}{\beta_{1,i}}\log_2\left(\frac{\mathbb{E}\Big[(\frac{1}{\rho}+ P_{1,i}|h_{1,i}|^2)^{\frac{2 \beta_{1,i}}{M}}\Big]}{\mathbb{E}\Big[(\frac{1}{\rho}+ 2 P_{1,i} |h_{1,i}|^2)^{\frac{\beta_{1,i}}{M}}\Big]}\right)\nonumber\\&+\frac{1}{\beta_{2,i}}\log_2\left(\frac{\mathbb{E}\Big[(1+ \frac{P_{2,i}|h_{2,i}|^2}{\frac{1}{\rho}+  P_{1,i}|h_{1,i}|^2})^{\frac{2 \beta_{2,i}}{M}}\Big]}{\mathbb{E}\Big[(\frac{1}{\rho}+ 2 P_{2,i}|h_{2,i}|^2)^{\frac{\beta_{2,i}}{M}}\Big]}\right)\Bigg).
\end{align}

\begin{align}
&\lim\limits_{\rho\rightarrow\infty}(E_c^{tot}-\tilde{E_c}^{tot})=\nonumber\\
 &\sum_{i=1}^{\frac{M}{2}}\!\Bigg(\frac{1}{\beta_{1,i}}\log_2\left(2^{-\frac{\beta_{1,i}}{M}} \mathbb{E}\Big[(  P_{1,i}|h_{1,i}|^2)^{\frac{\beta_{1,i}}{M}}\Big]\right)\nonumber\\
 &+\frac{1}{\beta_{2,i}}\log_2\left(\frac{\mathbb{E}\Big[\left(1+ \frac{P_{2,i}|h_{2,i}|^2}{ P_{1,i}|h_{1,i}|^2}\right)^{\frac{2 \beta_{2,i}}{M}}\Big]}{\mathbb{E}\Big[( 2 P_{2,i} |h_{2,i}|^2)^{\frac{\beta_{2,i}}{M}}\Big]}\right)\Bigg),\nonumber
\end{align}
which is a constant with respect to $\rho$.

Furthermore, to analyze
$\lim\limits_{\rho\rightarrow 0}(\frac{\partial (E_c^{tot}-\tilde{E_c}^{tot})}{\partial\rho})$ and $\lim\limits_{\rho\rightarrow\infty}(\frac{\partial (E_c^{tot}-\tilde{E_c}^{tot})}{\partial\rho})$, we start with $\frac{\partial E_c^{tot}}{\partial\rho}$ and $\frac{\partial \tilde{E_c}^{tot}}{\partial\rho}$.

\begin{align}
   &\frac{\partial E_c^{tot}}{\partial\rho}=\sum_{i=1}^{\frac{M}{2}}\bigg(\frac{\partial E_c^{1,i}}{\partial\rho}+\frac{\partial E_c^{2,i}}{\partial\rho}\bigg)\nonumber\\
   &=\sum_{i=1}^{\frac{M}{2}}\bigg(\frac{2 P_{1,i}}{M \ln2} \frac{\mathbb{E}\Big[|h_{1,i}|^2 (1+\rho P_{1,i} |h_{1,i}|^2)^{\frac{2 \beta_{1,i}}{M}-1}\Big]}{\mathbb{E}\Big[(1+\rho P_{1,i} |h_{1,i}|^2)^{\frac{2 \beta_{1,i}}{M}}\Big]}\nonumber\\&\!\!+\frac{2 P_{2,i}}{M\ln2}\frac{\mathbb{E}\Big[\frac{ |h_{2,i}|^2}{(1+\rho P_{1,i} |h_{1,i}|^2)^2}(1+ \frac{\rho P_{2,i} |h_{2,i}|^2}{1+  \rho P_{1,i} |h_{1,i}|^2})^{\frac{2 \beta_{2,i}}{M}-1}\Big]}{\mathbb{E}\Big[(1+ \frac{\rho P_{2,i} |h_{2,i}|^2}{1+  \rho P_{1,i} |h_{1,i}|^2})^{\frac{2 \beta_{2,i}}{M}}\Big]}\bigg),
\end{align}
where (.)' a first derivative with respect to $\rho$. 
Then,
$ \lim\limits_{\rho\rightarrow 0}\left(\frac{\partial E_c^{tot}}{\partial\rho}\right)=\sum_{i=1}^{\frac{M}{2}}\left(\frac{2 P_{1,i}}{M \ln2}\mathbb{E}[|h_{1,i}|^2]+\frac{2 P_{2,i}}{M\ln2}\mathbb{E}[|h_{2,i}|^2\right).$

\begin{align}
&\lim\limits_{\rho\rightarrow\infty}\left(\frac{\partial E_c^{tot}}{\partial\rho}\right)=\lim\limits_{\rho\rightarrow\infty}\bigg(\sum_{i=1}^{\frac{M}{2}}\bigg(\frac{2}{M \ln2 \rho}\nonumber\\
& +\frac{2 P_{2,i}}{M\ln2 \rho^2}\frac{\mathbb{E}\Big[\frac{ |h_{2,i}|^2}{(P_{1,i} |h_{1,i}|^2)^2}\left(1+ \frac{P_{2,i} |h_{2,i}|^2}{P_{1,i} |h_{1,i}|^2}\right)^{\frac{2 \beta_{2,i}}{M}-1}\Big]}{\mathbb{E}\Big[\left(1+ \frac{P_{2,i} |h_{2,i}|^2}{P_{1,i} |h_{1,i}|^2}\right)^{\frac{2 \beta_{2,i}}{M}}\Big]}\bigg)\bigg)=0.\nonumber
\end{align}
Similarly,
\begin{align}
   &\frac{\partial \tilde{E_c}^{tot}}{\partial\rho}=\sum_{i=1}^{\frac{M}{2}}\left(\frac{\partial \widetilde{E}_c^{1}}{\partial\rho}+\frac{\partial \widetilde{E}_c^{2}}{\partial\rho}\right),\nonumber\\
   &=\sum_{i=1}^{\frac{M}{2}}\bigg(\frac{1}{M \ln2}\frac{\mathbb{E}\Big[2 P_{1,i}|h_{1,i}|^2 (1+2 \rho P_{1,i} |h_{1,i}|^2)^{\frac{ \beta_{1,i}}{M}-1}\Big]}{\mathbb{E}\Big[(1+2 \rho P_{1,i}|h_{1,i}|^2)^{\frac{ \beta_{1,i}}{M}}\Big]}\nonumber\\&+\frac{1}{M \ln2}\frac{\mathbb{E}\Big[2 P_{2,i}|h_{2,i}|^2 (1+2\rho P_{2,i} |h_{2,i}|^2)^{\frac{\beta_{2,i}}{M}-1}\Big]}{\mathbb{E}\Big[(1+2 \rho P_{2,i} |h_{2,i}|^2)^{\frac{ \beta_{2,i}}{M}}\Big]}\bigg).
\end{align}
Then we have that,
$ \lim\limits_{\rho\rightarrow 0}\left(\frac{\partial \tilde{E_c}^{tot}}{\partial\rho}\right)=\sum_{i=1}^{\frac{M}{2}}\left(\frac{2 P_{1,i}}{M \ln2}\mathbb{E}[|h_{1,i}|^2]+\frac{2 P_{2,i}}{M\ln2}\mathbb{E}[|h_{2,i}|^2\right)$,
and 
 $\lim\limits_{\rho\rightarrow \infty}\left(\frac{\partial \tilde{E_c}^{tot}}{\partial\rho}\right)=\lim\limits_{\rho\rightarrow \infty}\left(\sum_{i=1}^{\frac{M}{2}}\frac{1}{\rho M \ln2}+\frac{1}{\rho M\ln2}\right)=0$.
So that, 
$\lim\limits_{\rho\rightarrow0}\left(\frac{\partial (E_c^{tot}-\tilde{E_c}^{tot})}{\partial\rho}\right)
=0.$

By following similar approach, we also get, 
$\lim\limits_{\rho\rightarrow\infty}\left(\frac{\partial (E_c^{tot}-\tilde{E_c}^{tot})}{\partial\rho}\right)=0.$
\bibliographystyle{IEEEtran}
\bibliography{IEEEabrv,biblio}
\balance

\end{document}